%% file: main.tex
\documentclass[10pt,twocolumn,letterpaper]{article}

\usepackage[pagenumbers]{cvpr} %

\input{preamble}

\definecolor{cvprblue}{rgb}{0.21,0.49,0.74}
\usepackage[pagebackref,breaklinks,colorlinks,allcolors=cvprblue]{hyperref}

\def\paperID{13997} %
\def\confName{CVPR}
\def\confYear{2025}

\title{Seeing Sound: Assembling Sounds from Visuals for Audio-to-Image Generation}

\author{Darius Petermann\thanks{The work conducted during author's internship at Netflix.}\\
Indiana University\\
Bloomington, IN\\
{\tt\small daripete@iu.edu}
\and
Mahdi M. Kalayeh\\
Netflix\\
Los Gatos, CA\\
{\tt\small mkalayeh@netflix.com}
}

\begin{document}
\maketitle

\begin{strip}
    \centering
    \vspace{-3em}
    \includegraphics[width=1.\linewidth]{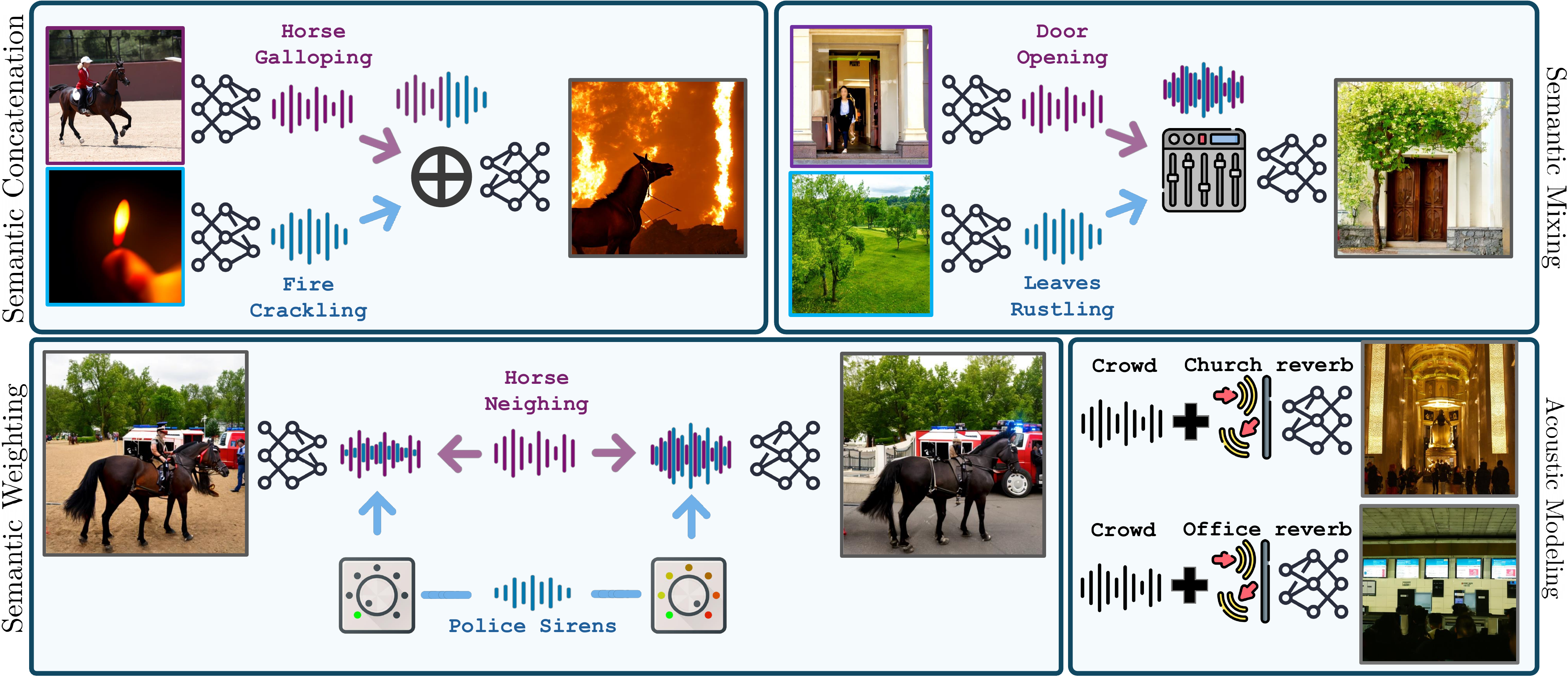}
    \captionof{figure}{
        Our audio-visual data modeling demonstrates versatile control for image generation through audio manipulations, including loudness calibration, audio mixing, and reverberations, showcasing our model's adaptability across a wide and unconstrained data domain.
    }
    \label{fig:main}
\end{strip}

\input{sec/0_abstract}

\input{sec/1_intro}
\input{sec/2_related}

\input{sec/3_method}

\input{sec/5_experiments}

\input{sec/6_conclusions}

{
    \small
    \bibliographystyle{ieeenat_fullname}
    \bibliography{main}
}
\input{sec/7_appendix}

\end{document}

%% file: preamble.tex
\usepackage{upgreek}
\usepackage{etoolbox,siunitx}
\usepackage{cuted}
\usepackage{pifont}
\usepackage{algorithm}
\usepackage{amsmath,amsfonts,amssymb,amsthm}
\usepackage[noend]{algpseudocode} 
\usepackage{afterpage}   %
\usepackage{enumitem}
\setlist{nosep}

\DeclareMathAlphabet{\pazocal}{OMS}{zplm}{m}{n}

\newcommand{\VLM}{\mathbb{V}}

\newcommand\indomain[1]{\textcolor[HTML]{005400}{#1}}
\newcommand\outdomain[1]{\textcolor[HTML]{820000}{#1}}

\algnewcommand{\LeftComment}[1]{\Statex \(\triangleright\) #1}

%% file: sec/0_abstract.tex
\begin{abstract}
Training audio-to-image generative models requires an abundance of diverse audio-visual pairs that are semantically aligned. Such data is almost always curated from in-the-wild videos, given the cross-modal semantic correspondence that is inherent to them.
In this work, we hypothesize that insisting on the absolute need for ground truth audio-visual correspondence, is not only unnecessary, but also leads to severe restrictions in scale, quality, and diversity of the data, ultimately impairing its use in the modern generative models.
That is, we propose a scalable image sonification framework where instances from a variety of high-quality yet disjoint uni-modal origins can be artificially paired through a retrieval process that is empowered by reasoning capabilities of modern vision-language models. To demonstrate the efficacy of this approach, we use our sonified images to train an audio-to-image generative model that performs competitively against state-of-the-art. Finally, through a series of ablation studies, we exhibit several intriguing auditory capabilities like semantic mixing and interpolation, loudness calibration and acoustic space modeling through reverberation that our model has implicitly developed to guide the image generation process. 
\end{abstract}

%% file: sec/1_intro.tex
\section{Introduction}
\label{sec:intro}
\textit{Can a single auditory scene correspond to multiple visual interpretations?} Consider the sound of a horse walking on concrete coupled with nearby women chattering and leaves rustling in the background. Seeking to visualize such a scene given its sonic counterpart is a non-trivial and inherently ambiguous task. While many semantics conveyed in audio (\textit{e.g.} horse, women, and leaves) have strong visual correspondences, others may not. What types of clothing are worn by the women? Is the rustling noise emitted from a tall tree or a small bush? What color is the horse's coat? Answering these questions involves semantics that are inherently unbound to any auditory quality, making it impossible to reason about them solely from sound. Training a model for such a cross-modal translation task is a challenging undertaking, and most of the existing work on this front has exploited in-the-wild videos for that purpose~\cite{lee2022sgsim,Zhou2017VisualTS,lee2022landscapes,li2022into-the-wild,yariv2023audiotoken,sung2023soundscene,biner2024sonicdiffusion},
banking on the synchronous audio which naturally accompanies the visuals. However, the audio excerpts extracted from such videos could lack sufficient quality and semantic correspondence, not to mention the limited diversity rooted in their meticulously curated domains. In this context, we argue that the reliance on in-the-wild data is overstated. We present a framework providing beyond-versatile sonification capabilities, allowing any image dataset to be sonified, thus expanding the potential applications across various domains.

In recent years, diffusion models have thrived through their exceptional generative capabilities \cite{nichol2022glide, saharia2022imagen, ramesh2022dalle2,chen2023pixartalpha}. 
Approaches such as~\cite{ramesh2022dalle2, rombach2021highresolution, saharia2022imagen} have demonstrated remarkable success in generating high-quality images, yet their reliance on textual descriptions alone may lead to certain limitations particularly as natural language can fail at accurately capturing the rich nature of a visual scene when conciseness is of essence.
That is, early efforts have started exploring audio as a plausible conditioning signal \cite{lee2022sgsim, sung2023soundscene,yariv2023audiotoken,biner2024sonicdiffusion} for several reasons. First, the \emph{temporal dynamics} bound to sound provides rich context and nuanced continuous flow of information. Transitions, rhythm, pace, and progression are all key nuances that are often lost in the discrete nature of textual representation. Second, the \emph{rich expressiveness} intrinsic to sound can efficiently convey the information more abundantly and succinctly in contrast to text. Emotions, intentions, and subtle nuances through variations in tone, pitch, and volume, in the context of dialog, can efficiently transmit contextual information.
Third, the \emph{natural multi-modal relation} between the two modalities makes audio a suitable mean to steer image generation. After all, sight and sound are deeply intertwined in human perception, and their co-occurrence in our everyday experiences is ubiquitous.

The challenge of aligning audio with image is not trivial however; existing methods have typically struggled with data quality and cross-modality correspondence.
Prior efforts in audio-driven image generation \cite{lee2022sgsim,sung2023soundscene,qin2023gluegen,yariv2023audiotoken,biner2024sonicdiffusion}
have paved the path for exploring novel generation avenues,
yet their reliance on video assets substantially limits their diversity, domain coverage, and generalization capabilities. 
More importantly, the quality and consistency of the training pairs remains largely affected by the noisy and low-quality nature of the original video assets.

In this paper, 
we introduce a novel image sonification scheme which yields high-quality audio-visual pairs by leveraging a plethora of diverse disjoint uni-modal image and audio datasets. 
Our robust sonic generation process ensures high perceptual quality for individual modalities and superior cross-modal semantic correspondence 
through a modular, and flexible retrieval method. By leveraging pre-trained vision-language models (VLMs)~\cite{liu2023llava, wang2023cogvlm} and multi-modal latent representations~\cite{radford2021clip, clap2023wu}, we enhance the image-to-audio retrieval process which ultimately results in improved data quality and relevance. Our approach enables domain-specific adaptation of pre-trained text-to-image diffusion models to the audio-to-image generation task, maintaining domain consistency and ultimately enhancing model performance. 
Our contributions are as follows:

\begin{itemize}
\item We propose a recipe for building large-scale audio-visual pairs of high correspondence from uni-modal image and audio datasets through a cross-modal retrieval process that enables arbitrary image datasets to be sonified.
\item We adapt pre-trained text-to-image models for audio-to-image generation, and show competitive performance against state-of-the-art on multiple complementary metrics across five different benchmarks. 
\item We conduct a series of ablation studies to analyze various auditory properties emerged from our framework as shown in Figure~\ref{fig:main}.
\item We will open-source the code and model weights, in addition to our audio-visual dataset consisting of approximately 1M images, paired with textual sounding concepts and corresponding audio counterparts.
\end{itemize}

%% file: sec/2_related.tex
\section{Related Work}
\label{sec:related}
\textbf{Modeling Paradigms } Early research in the image generation primarily involved the application of the \emph{generative adversarial networks} (GAN) \cite{goodfellow2014gan} to synthesize realistic images from simple text often in form of category labels.
Over the years, these approaches improved substantially in image resolution, editing capabilities, and inference efficiency~\cite{reed2016gan, zhang2017gan, Xu2017AttnGANFT, Brock2018LargeSG, zhu2019gan, zhang2019stackgan, karras2019style}. With the recent progress in diffusion-based models \cite{ho2020denoising, dhariwal2021diffusion, rombach2021highresolution, nichol2022glide, ramesh2022dalle2, saharia2022imagen, Peebles2022DiT, chen2023pixartalpha}, the field of text-to-image generation has been revolutionized. These models not only considerably surpass GAN-based approaches in generation quality and realism but also provide more stable optimization while enabling far more complex conditioning functionalities. It is worth emphasizing that many of these improvements were achieved in part thanks to advancements in other areas including self-supervised representation learning ~\cite{radford2021clip}, neural architectures~\cite{dosovitskiy2021an, Peebles2022DiT} and large language models~\cite{raffel2020exploring, touvron2023llamaopenefficientfoundation}, with modern LLMs enabling a remarkable understanding of very complex natural text compared to their predecessors~\cite{devlin2018bert}.

Motivated by the success of text-to-image models, recent literature shows increasing interest in exploiting other modalities, such as audio, to drive or enhance the process of pixel generation. Most of the work in this area relies on some kind of pre-trained joint multi-modal latent space ~\cite{radford2021clip, guzhov2022audioclip, clap2023wu} to enable the transition across modality boundaries. With that in place, one can then adapt text-to-image architecture to accept novel modalities ~\cite{lee2022sgsim, sung2023soundscene, qin2023gluegen, girdhar2023imagebind}.

Recently, a handful of audio-only methods have also emerged. Audiotoken \cite{yariv2023audiotoken} and SonicDiffusion~\cite{biner2024sonicdiffusion} expand the capabilities of LDMs~\cite{rombach2021highresolution} beyond text-to-image generation and towards audio. Specifically, Audiotoken \cite{yariv2023audiotoken} leverages a pre-trained text-to-image LDM together with a pre-trained audio embedding model and learns an adaptation layer which maps between their outputs and inputs. This enables the use of a dedicated audio token to condition the generation process without any text input. In contrast, SonicDiffusion~\cite{biner2024sonicdiffusion} first learns a joint audio-text space through modifications in cross-attention layers of \cite{rombach2021highresolution} which ultimately facilitates the intake of audio and text modalities for image generation. At inference, with text input being optional, their model can operate in an audio-only fashion. Our work is closely related to~\cite{yariv2023audiotoken, biner2024sonicdiffusion} as we also operate solely from audio and adapt a pre-trained text-to-image diffusion model~\cite{chen2023pixartalpha} to do audio-to-image generation.

\noindent\textbf{Data Paradigms }Finding large-scale and semantically diverse audio-visual data with strong cross-modal correspondence is not trivial. The literature has been focused on curating audio-visual data from in-the-wild videos given that they naturally ensure synchronized modalities. In doing so, most of the prior works have overemphasized on cross-modal correspondence, at the cost of severely restrained semantic diversity and scale~\cite{greatest_hits, Zhou2017VisualTS, lee2022landscapes, li2022into-the-wild}.
An exception is VGGSound~\cite{chen2020vggsound} which was introduced with 200K videos across 309 audio classes to partially address the shortcomings (\textit{e.g.} noisy audio, low-resolution video, low cross-modality correspondence etc.)
of the widely popular Audioset~\cite{gemmeke2017audioset} that hindered its effective use for audio-visual representation learning.
Such a data landscape forces compromises on scale, quality, and diversity when training modern generative architectures.

In this work, we advocate that the reliance on in-the-wild data in order to get semantically correspondent audio-visual pairs is exaggerated and unnecessary. We will describe a scalable framework to curate semantically aligned audio-visual pairs from a pool of diverse and high quality yet disjoint uni-modal datasets. 
We demonstrate the effectiveness of our approach by training a single audio-to-image diffusion model and evaluating it across multiple out-of-domain datasets~\cite{greatest_hits, Zhou2017VisualTS, lee2022landscapes, li2022into-the-wild} from landscapes and nature scenery to concepts like fireworks, and human activities.

%% file: sec/3_method.tex
\begin{figure*}
    \centering
        \includegraphics[width=1.0\textwidth]{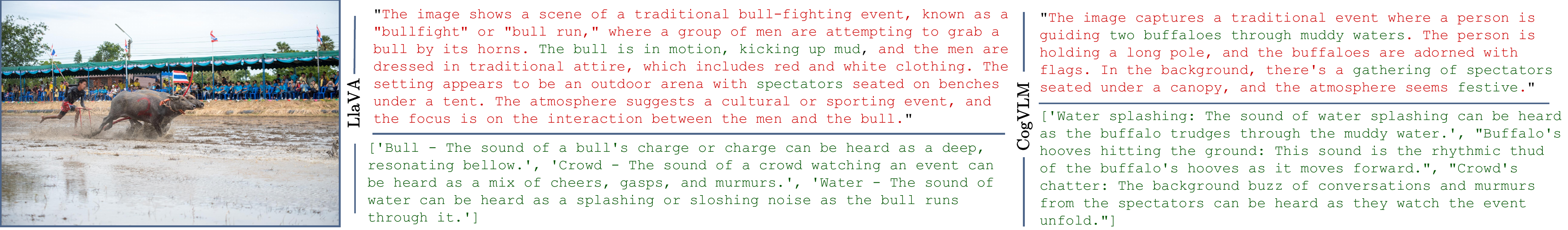}
    \vspace*{-6mm}
    \caption{Visually-aligned (up) vs. sonically-aligned (down) image descriptions using \emph{LLaVA} \cite{liu2023llava} and CogVLM \cite{wang2023cogvlm}. Each description is obtained via its respective prompt: ``\textit{Provide a short and concise description of the following image.}'' and ``\textit{As a numbered list, provide one to up to three sound(s) associated with prominent objects visible and present in the image. Provide the objects followed by their associated sound}''. A large portion of the comprehensive description (\textcolor{red}{red}) does not pertain to acoustics properties whereas a few limited keywords do (\textcolor{green}{green}). Through handcrafted prompting we manage to obtain sonically-aligned and acoustically relevant descriptors.}
    \label{fig:prompts}
\vspace{-3mm}
\end{figure*}
\section{Method}
\label{sec:method}
Building a dataset consisting of audio-visual pairs with strong cross-modal correspondence has been traditionally tackled by retrieving short snippets directly from in-the-wild videos \cite{greatest_hits, Zhou2017VisualTS, gemmeke2017audioset,chen2020vggsound,lee2022landscapes}. 
In doing so, the inherent semantic mismatch between the audio and image domains brings several significant limitations. While the former conveys temporal information, the latter merely represents a fragment of it. This discrepancy naturally leads to the question of which visual fragment from the video to settle on. Another crucial aspect is data domain and quality of instances. In-the-wild data is noisy by nature and therefore a strong audio-visual correspondence is challenging to obtain. To partially overcome this problem, existing methods restrain their domains to sounding concepts 
with strong correspondence, severely limiting the capabilities of the models to handle diverse and out-of-domain queries at inference. Finally, there are very few curated large-scale and publicly available audio-visual datasets, the most popular one perhaps being Audioset~\cite{gemmeke2017audioset}. While very large in size, the instances in Audioset~\cite{gemmeke2017audioset} remain of relatively low quality for both modalities.
While existing methods put considerable efforts in addressing the aforementioned limitations, 
we argue that the need for in-the-wild data in the context of audio-to-image generation is overstated. Precisely, being bound to ``real'' audio-image pairs imposes a substantial and unnecessary constraint on the modeling task while excessively limiting the domains in which the model could operate. Moreover the natural audio-visual correspondences subsisting across in-the-wild pairs are, in essence, minimal due to the semantic gap across modalities, which further undermines the need for such a data in the first place. In this work, we therefore build our audio-visual pairs via a modular retrieval process, resulting in high-quality samples with strong cross-modality correspondence. 
In the following sections, we go over the process of obtaining plausible, accurate, and strongly correspondent sonic representation for high-resolution images.

\subsection{Notations}
We denote the $n^{th}$ audio snippet with $a_n \in \mathbb{R}^{T}$, where $T$ is the temporal resolution. The pool of our audio excerpts is therefore denoted by $\mathcal{A}=\{a_{n}| n\in[1\cdots N]\}$, where $N$ represents the total number of audio snippets. 
Similarly, the pool of $M$ images is denoted by $\mathcal{I}=\{i_{m}| m\in[1\cdots M]\}$. Note that $N \neq M$ since they are comprised of disjoint unimodal datasets. Let $\VLM$ be representing a VLM that we utilize to extract sounding concepts from images. That is, $\VLM$ maps the input image $i_m$ to a set of textual outputs $ \mathcal{S}_m = \VLM(i_m)$. 
We use pre-trained modality encoders $\mathcal{E}_{a}$ and $\mathcal{E}_{t}$, to respectively transform instance from $\mathcal{A}$ and $\mathcal{S}$ to a joint latent space where cosine similarity is semantically meaningful.

\subsection{Extracting Sounding Concepts}

Prior works have harnessed the power of state-of-the-art vision-language models, such as LLaVA \cite{liu2023llava}, towards complex reasoning and labeling tasks. In this work, we do not seek to acquire a comprehensive textual description of an image but rather one that specifically pertains to the \emph{sonic qualities} possibly portrayed in the image (\textit{i.e.} sounding descriptors). For instance, provided with the visual of a brown dog playing fetch near a river at sunset, we would ideally like to extract sounding concepts such as ``dog'' and ``river'' while putting aside the remaining of the description such as ``sunset'' and ``brown'' as these terms don't pertain to any acoustic qualities.
Figure~\ref{fig:prompts} exemplifies the process behind instructing a VLM with an appropriate prompt, leading to an effective \emph{sounding concept} extraction.
Throughout a comprehensive empirical assessment of various VLMs and prompting methods, we found that \emph{LLaVA}~\cite{liu2023llava} and \emph{CogVLM}~\cite{wang2023cogvlm} were both generating satisfactory responses albeit with differences in retrieved concepts. With that in mind, we opt to incorporate both models as part of our sonification pipeline,
adding more diversity to our pool of sounding concepts and introducing an augmentation mechanism in which a single image is bound to more than one plausible acoustic scenes. Our final prompt (ref. Figure~\ref{fig:prompts}) is the result of careful engineering trials. For instance we found that VLMs often make assumptions about sounding objects that may be off-camera, which could be detrimental to our task. To address this, adding the clarification \emph{visible and present in the image} significantly improved the retrieval quality. We share more examples of the extracted sounding concepts from diverse set of images in Appendix \ref{subsec:appendix_datasetoverview}.

\subsection{From Sounding Concepts to Audio Excerpts}
\label{subsec:concepts}
We have established that modern VLMs when adequately prompted, are capable of extracting sounding concepts as text (ref. Figure~\ref{fig:prompts}). After obtaining accurate descriptive texts for each of the sounding concepts
the next challenge is to retrieve their sonic counterparts while minimizing the loss of information in the modality translation process. Consider the following sounding concept: 

\emph{dog: sound of a dog barking while running in the grass}.
Here, we not only need to account for the dog barks but for the rustling grass produced by the dog steps as well. To do so, we exploit the cross-modal pre-trained latent space of CLAP \cite{clap2023wu} to conduct text-to-audio retrieval. Specifically, we utilize its audio encoder to generate an embedding for every audio snippet resulting in $\mathcal{E}_{a}(\mathcal{A})\in \mathbb{R}^{D}$, the equivalent of $\mathcal{A} \in \mathbb{R}^{T}$ but in the $D$-dimensional latent space of CLAP \cite{clap2023wu}. Similarly, its text encoder, denoted by $\mathcal{E}_{t}$, allows us to transform each textual sounding concept $s \in \mathcal{S}$ to its corresponding latent embedding $\mathcal{E}_{t}(s) \in \mathbb{R}^{D}$. With that, retrieving a semantically-aligned audio snippet from $\mathcal{A}$ for a given sounding concept $s$ boils down to a random process with $\mathbb{P}(a|s)\propto\mathcal{E}_{t}(s)^\intercal\mathcal{E}_{a}(a)$. However, this formulation is sub-optimal for our task given that outliers are present in the Euclidean vector space. Consider two audio excerpts, while $a^{dog}$ solely conveys the sound of a ``dog barking'', $a^{dog+}$ additionally indicates other sources such as ``people speaking'', or ``leaves rustling'' as well.
It is reasonable to assume that for a large portion of images showing a dog, the VLM yields some sounding concept $s^{dog}$ that is textually quite close to ``dog barking''. In such a scenario, we have consistently observed that $\mathcal{E}_{t}(s^{dog})^\intercal\mathcal{E}_{a}(a^{dog}) \gg  \mathcal{E}_{t}(s^{dog})^\intercal\mathcal{E}_{a}(a^{dog+})$ leading to over-sampling of $a^{dog}$ upon termination of the retrieval process at scale. To ameliorate this, we apply signed square root (SSR) on top of cross-modal similarity scores to suppress outliers and reduce the overall variance among the top matches. This would lead to a larger and more diverse pool of audio excerpts to get paired with our images. To ensure high cross-modal correspondence, and efficient sampling ($N$ is  about 500K), we dynamically estimate a threshold to lower bound the similarity scores.
We use average score among top-$k$ matches per query. This approach would yield a smaller pool of audio snippets when similarity score drastically reduce as $k$ grows (\textit{i.e.} few very good matches followed by many mediocre ones). With the same logic, the pool gets larger when diversity of similarity scores is rather low among the top matches (\textit{i.e.} many equally good matches). Algorithm~\ref{algo:retrieval} details our retrieval process for a single sounding concept. In practice, we implement and execute in parallel across all the sounding concepts associated with all of our images.

\begin{algorithm}
\small
\caption{Retrieving a relevant audio snippet for a textual sounding concepts $s$}
\label{algo:retrieval}
\begin{algorithmic}[1]

\Procedure{\texttt{GetMatchedAudio}}{$s, \mathcal{A}, \mathcal{E}$}
    
    \State $\texttt{p}\gets\mathcal{E}_{a}(\mathcal{A})$ \Comment{encode audio excerpts}    
    \State $\texttt{q}\gets\mathcal{E}_{t}(s)$  \Comment{encode query sounding concept}        
    
    \State $\texttt{t} \gets \texttt{CosSim(q,p)}$ \Comment{get similarity scores}
    \State $\texttt{t}\gets \texttt{sign(t)}\cdot \texttt{sqrt(abs(t))}$ \Comment{apply SSR}      
    \State $\texttt{lb} \gets \texttt{mean(topk(t))}$ \Comment{get min eligible score}            
    \State $\texttt{I} \gets \texttt{argwhere(t > lb)}$ \Comment{find top matches}             
    \State $\texttt{i} \gets \texttt{choice(I)}$ \Comment{randomly pick a match}            
    \State \textbf{return} $\mathcal{A}\texttt{[i]}$ \Comment{return sampled audio snippet}
\EndProcedure
\end{algorithmic}
\end{algorithm}

In order to obtain the final audio counterpart $a_m$ for a given image $i_m$, the sounding concepts retrieved from the Algorithm ~\ref{algo:retrieval} first have their loudness individually normalized using a value $\gamma_m$ sampled from a uniformly-distributed decibel-LUFS \cite{grimm2010lufs} range, before being linearly summed up in the time-domain to obtain $a_m$. We provide more details on the nature of LUFS normalization and its calculation in Appendix \ref{subsec:appendix_lufs}.

\subsection{Audio Representations}
\label{sec:audio_representation}
Raw audio waveforms are difficult to deal with within a learning framework mainly given their large sample size, redundancy, and lack of semantic abstraction. These are precisely the aspects which latent space of pre-trained audio models are known to accommodate for rather adequately. While seeking semantically meaningful representations with sufficient temporal resolution, several considerations have to be accounted for since the properties carried by these representations will define the extent of audio-driven control that the generative model would ultimately enjoy. In this view, we aim for \emph{loudness retention} and \emph{multi-source disentanglement}. Specifically, how much of the signal amplitude is retained as part of the audio embeddings and whether these representation are capable of concurrently encoding multiple audio sources. The former is useful for weighting the semantics as part of the generation process (\textit{e.g.} the louder in the audio domain, the more prominent in the visual domain), while the latter is necessary as we rarely have audio excerpts during training that are truly single-source. More importantly, this characteristic would unlock exciting inference-time semantic functionalities like those illustrated in Figure~\ref{fig:main}. 

With that in mind, we consider the Audio Spectrogram Transformer (AST)~\cite{gong2021ast}, an audio classifier, to  a suitable and meaningful representation. While multi-source disentanglement is a given considering the upstream multi-label classification pre-training of AST~\cite{gong2021ast}, it is not a priori clear how well the notion of signal loudness is retained within its embeddings. We hence designed a study to explore this and concluded that the embeddings do carry such characteristic to a decent extent. Appendix \ref{subsec:appendix_astloudness} contains details of the aforementioned study. All the experiments reported in this work are hence using AST~\cite{gong2021ast} to convert $a_m$ to a latent representation in form of vector time-series.

%% file: sec/5_experiments.tex
\section{Experiments}
\begin{figure*}
    \centering
        \includegraphics[width=\textwidth]{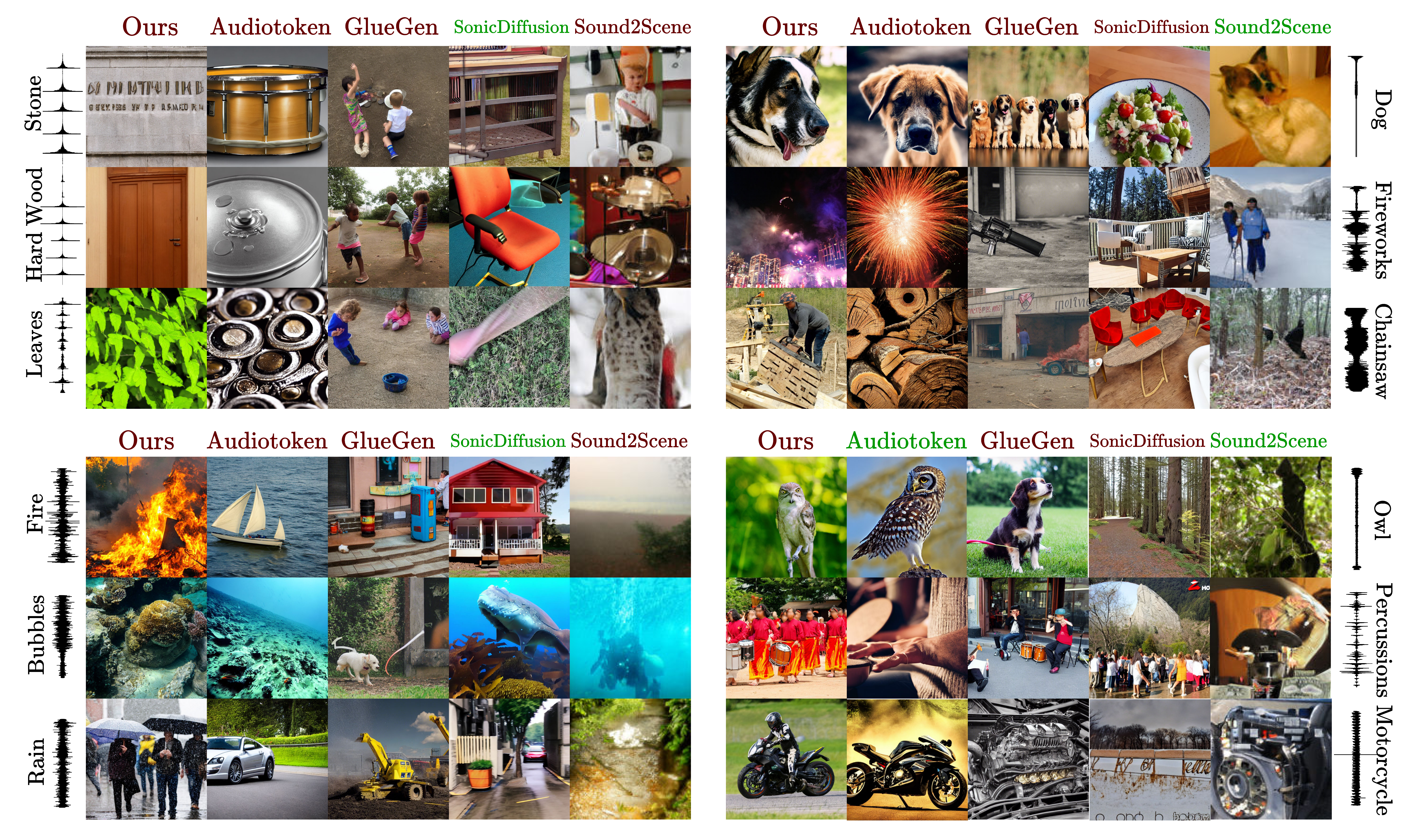}
    \vspace{-4mm}
    \caption{Qualitative comparison of state-of-the-art audio-to-image generative models on four different datasets; Greatest Hits (top left), VEGAS (top right), Landscapes + ITW (bottom left), VGGSound (bottom right). Models are highlighted in \indomain{green} if they are evaluated on \indomain{in-sample} data and in \outdomain{red} if evaluated on \outdomain{out-of-sample} data. Our model consistently performs well on out-of-sample data, with results that are on par and most often exceeding those of in-sample models.}
    \label{fig:domains}
\vspace{-3mm}
\end{figure*}

\subsection{Datasets}
\label{subsec:dataset}
We build our audio-visual pairs, detailed in Sec.~\ref{sec:method}, on top of several well-established uni-modal audio and vision datasets. 
Starting with audio, our pool of recordings is assembled from three large-scale sources, namely Audioset~\cite{gemmeke2017audioset}, FSD50K~\cite{fonseca2022FSD50K}, and BBC SFX\footnote{BBC sound effects library \url{https://sound-effects.bbcrewind.co.uk/}}. Audioset~\cite{gemmeke2017audioset} stems from a large collection of Youtube videos and contains more than 500 sound events. Each instance is an $\sim$10 second long video clip from which the audio track has been extracted. We use the balanced set, which provides us with at least 59 examples for each sound event yielding $\sim$22K recordings and totaling 61.3 hours of audio data. FSD50K~\cite{fonseca2022FSD50K} consists of an open dataset of human-labeled sound events containing Freesound clips distributed across 200 classes drawn from the AudioSet~\cite{gemmeke2017audioset} ontology. The clip lengths range from 1 up to 30 seconds and totals 108.3 hours of multi-labeled audio. BBC SFX is a large collections of foleys, sound effects, and nature recordings made for specific program-making purposes. The dataset has been curated and annotated by the BBC and contains more than 16K recordings with a total runtime of more than 460 hours. All three audio datasets combined results in about 0.5M audio excerpts, all 5-second long, across more than 500 audio categories. 
The visual portion of our data consists of 1M randomly selected images from SAM~\cite{kirillov2023segany}. Originally designed for large-scale image segmentation, SAM has recently been used across numerous vision tasks \cite{zhang2023samexplore,chen2023pixartalpha} given its scale, semantic diversity and open-source nature. 
To prevent non-sonic images, like icons, logos, and symbols, from impairing the training process, we use VLMs to detect and filter them out. For more on data curation and standardization, readers can refer to Appendix \ref{subsec:appendix_filtering}.

\subsection{Evaluation Metrics}
\label{subsec:eval}

Following recent literature, we evaluate different audio-to-image generative models using the following metrics.

\noindent\textbf{Audio-Image Similarity (AIS)} aims at evaluating the alignment between a generated image and its audio counterpart, namely by computing cosine similarity between latent embeddings associated with the two modality instances.
In accordance with~\cite{yariv2023audiotoken, biner2024sonicdiffusion}, we also employ Wav2CLIP~\cite{wu2022wav2clip} for this purpose. \textbf{Image-Image Similarity (IIS)} is the image analog to what AIS is for audio and measures the semantic similarity between generated and ground-truth\footnote{real image corresponding to audio prompt} images. We utilize the visual tower of CLIP~\cite{hessel2021clipscore} to obtain latent representations from images when implementing IIS. Finally, we adopt \textbf{Fr\'echet Inception Distance (FID)}~\cite{Heusel2017GANsTB} 
to quantify the distance between distribution of the generated and ground-truth images. Unlike AIS, FID does not directly take audio-to-image alignment into account, rather it implicitly measures perceptual quality and diversity at a distribution level.

\subsection{Training Details}
\label{subsec:train_details}
We use the pre-trained variational autoencoder (VAE) from LDM~\cite{rombach2021highresolution} and AST~\cite{gong2021ast} to respectively pre-compute the latent representation of training images and their corresponding audio excerpts.
We initialize\footnote{layers that map input audio embedding to latent space of diffusion model are randomly initialized} our model with a checkpoint from pre-trained PixArt-$\alpha$~\cite{chen2023pixartalpha} at 512$\times$512 and train for about 0.5M steps on 8$\times$A100 GPUs with a total batch size of 640. We use AdamW~\cite{Loshchilov2017DecoupledWD} with learning rate of 3.16e-5 and weight decay is set to 3e-2.

\subsection{Results}
\label{sec:results}

\begin{table*}[t]
\scriptsize
\centering
  \sisetup{
    detect-weight, %
    mode=text, %
    table-number-alignment=center
    }
\begin{tabular*}{\textwidth}{@{\extracolsep{\fill}} l 
    S[table-format=3.1] S[table-format=1.2] S[table-format=1.3]
    S[table-format=3.1] S[table-format=1.2] S[table-format=1.3]
    S[table-format=3.1] S[table-format=1.2] S[table-format=1.3]
    S[table-format=3.1] S[table-format=1.2] S[table-format=1.3]}
    \toprule
    & \multicolumn{3}{c}{Greatest Hits} & \multicolumn{3}{c}{Landscapes + ITW} & \multicolumn{3}{c}{VEGAS} & \multicolumn{3}{c}{VGGSound} \\
    \cmidrule(lr){2-4} \cmidrule(lr){5-7} \cmidrule(lr){8-10} \cmidrule(lr){11-13}
    & {FID $\downarrow$} & {IIS $\uparrow$} & {AIS $\uparrow$} & {FID $\downarrow$} & {IIS $\uparrow$} & {AIS $\uparrow$} & {FID $\downarrow$} & {IIS $\uparrow$} & {AIS $\uparrow$} & {FID $\downarrow$} & {IIS $\uparrow$} & {AIS $\uparrow$} \\
    \midrule
    Sound2Scene~\cite{sung2023soundscene} & \cellcolor{red!20} 184.2 & \cellcolor{red!20} \textbf{0.68} & \cellcolor{red!20} 0.049 & \cellcolor{red!20} 109.0 & \cellcolor{red!20} 0.71 & \cellcolor{red!20} 0.053 & \cellcolor{green!20} \textbf{94.0} & \cellcolor{green!20} \textbf{0.62} & \cellcolor{green!20} 0.048 & \cellcolor{green!20} \textbf{85.8} & \cellcolor{green!20} \textbf{0.60} & \cellcolor{green!20} 0.037 \\
    GlueGen~\cite{qin2023gluegen} & \cellcolor{red!20} 192.0 & \cellcolor{red!20} 0.58 & \cellcolor{red!20} 0.022 & \cellcolor{red!20} 234.5 & \cellcolor{red!20} 0.54 & \cellcolor{red!20} 0.051 & \cellcolor{red!20} 155.1 & \cellcolor{red!20} 0.51 & \cellcolor{red!20} 0.029 & \cellcolor{red!20} 140.0 & \cellcolor{red!20} 0.49 & \cellcolor{red!20} 0.030 \\
    Audiotoken~\cite{yariv2023audiotoken} & \cellcolor{red!20} 190.6 & \cellcolor{red!20} 0.60 & \cellcolor{red!20} \textbf{0.053} & \cellcolor{red!20} 132.9 & \cellcolor{red!20} 0.62 & \cellcolor{red!20} 0.057 & \cellcolor{red!20} 125.8 & \cellcolor{red!20} 0.57 & \cellcolor{red!20} 0.045 & \cellcolor{green!20} 115.7 & \cellcolor{green!20} 0.55 & \cellcolor{green!20} 0.034 \\
    SonicDiffusion~\cite{biner2024sonicdiffusion} & \cellcolor{green!20} \textbf{87.3} & \cellcolor{green!20} 0.72 & \cellcolor{green!20} 0.029 & \cellcolor{green!20} \textbf{79.9} & \cellcolor{green!20} \textbf{0.75} & \cellcolor{green!20} \textbf{0.064} & \cellcolor{red!20} 126.6 & \cellcolor{red!20} 0.48 & \cellcolor{red!20} 0.038 & \cellcolor{red!20} 120.4 & \cellcolor{red!20} 0.47 & \cellcolor{red!20} 0.038 \\
    \midrule
    Ours & \cellcolor{red!20} 228.4 & \cellcolor{red!20} 0.61 & \cellcolor{red!20} \textbf{0.053} & \cellcolor{red!20} 98.5 & \cellcolor{red!20} 0.67 & \cellcolor{red!20} 0.057 & \cellcolor{red!20} 120.3 & \cellcolor{red!20} 0.54 & \cellcolor{red!20} \textbf{0.051} & \cellcolor{red!20} 103.5 & \cellcolor{red!20} 0.51 & \cellcolor{red!20} \textbf{0.043} \\
    \bottomrule
\end{tabular*}
    \vspace{-2mm}
    \caption{Quantitative comparison with state-of-the-art audio-to-image generative models. The cells marked in \textcolor{red!70}{red} show that the dataset (column) is out-of-sample with respect to the model (row), whereas a \textcolor{green}{green} cell means in-sample evaluation.}
    \label{tab:metrics}
\vspace{-2mm}
\end{table*}

We benchmark various models on the Greatest Hits~\cite{greatest_hits}, Landscapes~\cite{lee2022landscapes}, Into The Wild (ITW)~\cite{li2022into-the-wild}, VEGAS~\cite{Zhou2017VisualTS}, and VGGSound~\cite{chen2020vggsound}. While this is not the case for our model, we note that a number of these datasets have in fact been used to train some of the prior models, giving them an advantage in our evaluation setup as the test instances were previously observed.
For VGGSound~\cite{chen2020vggsound}, we sample approximately 1K instances from the original test set in a class-balanced fashion following~\cite{sung2023soundscene}. For other benchmarks, considering their rather small size, we utilize all the available instances across different splits, when given. Table \ref{tab:metrics} quantitatively compares our approach against state-of-the-art audio-to-image generative models. Here we can make a few observations: 

\noindent\textbf{First}, in most cases (8 out of 12), the top performance has been achieved by the the model variants that are being evaluated in-sample. This is expected since the test samples have been observed already and even over-fitting on training data could lead to decent metrics for the corresponding models. In Appendix \ref{subsec:appendix_overfitting}, we share evidence that indeed some of these models have memorized their training data. \textbf{Second}, if we only consider variants that operate out-of-sample (\textit{i.e.} highlighted in red), with the exception of Greatest Hits, our model achieves the best FID across all the benchmarks. Looking at IIS, our approach mostly achieves competitive (runner up) results except on VGGSound~\cite{chen2020vggsound} where we outperform other diffusion-based methods which like us do operate in an out-of-sample regime. \textbf{Third}, our approach excels in AIS, which measures alignment between generated images and audio prompts, achieving top results on 3 out of 4 benchmarks, even against in-sample variants. In summary, these results support our hypothesis that training audio-to-image generative models on top of sonified images is effective. This challenges the widely held assumption that ground-truth audio-visual correspondence is necessary for such tasks. For qualitative comparison, please refer to Figure~\ref{fig:domains}.

\subsection{Ablation Studies}

\begin{figure}
    \centering
        \includegraphics[width=1.0\linewidth]{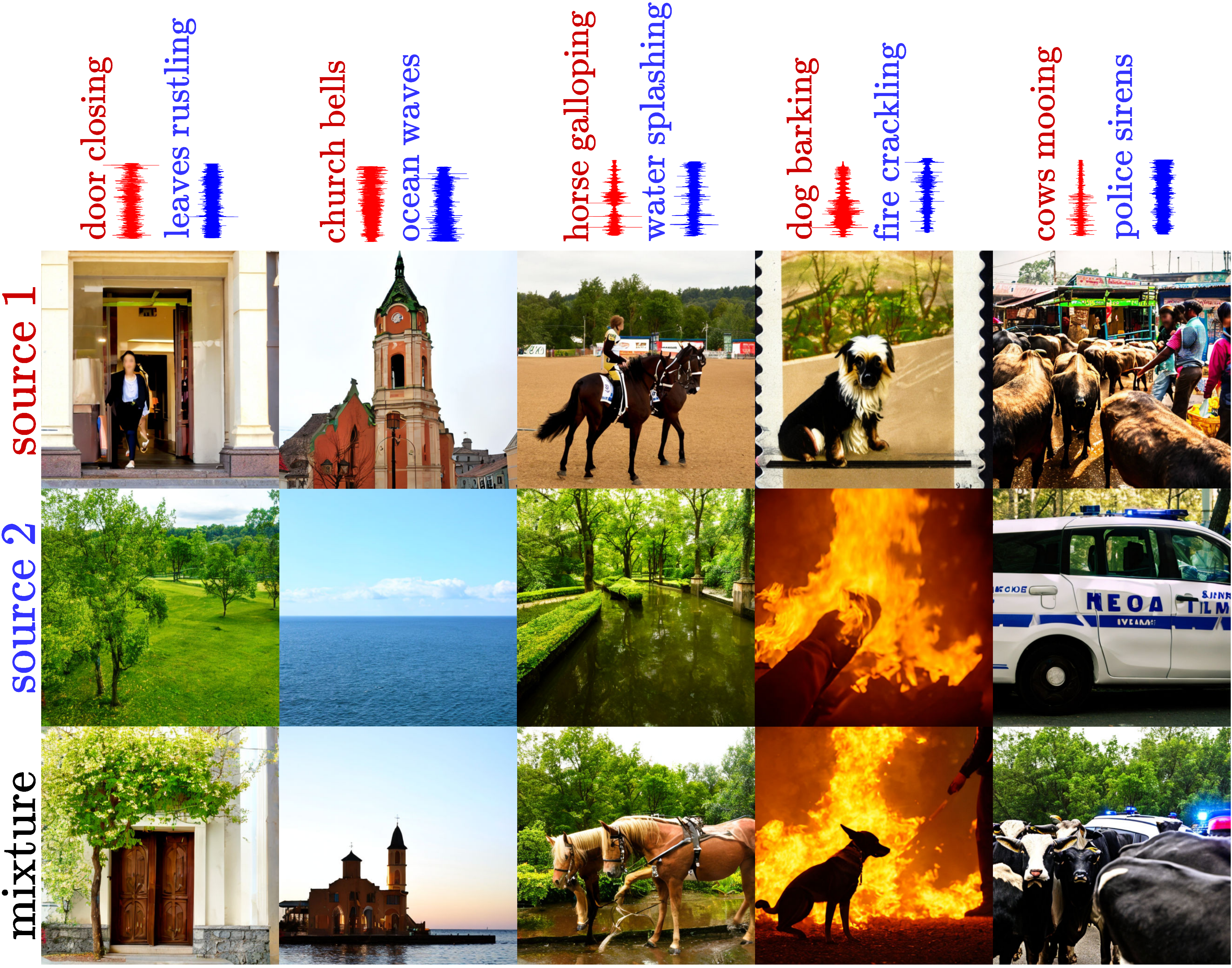}
        \vspace{-3mm}
    \caption{Examples of various semantics mixed together in the audio domain and their generated visual counterparts.}
    \label{fig:ablation_mix}
\vspace{-4mm}
\end{figure}

 \noindent\textbf{Semantic Mixing} We want to examine our model's ability to handle mixed audio inputs by combining multiple sources, each conveying different semantics. Our motivation is to observe how well the model generates images that reflect the combined semantics. This will not only demonstrate the model's capability to understand and integrate complex audio cues into coherent visual outputs, but also its ability to handle unique and novel semantic combinations. To do so, we randomly sample two excerpts from our pool of audio examples and mix them linearly in the time-domain before computing their AST~\cite{gong2021ast} embeddings. Figure~\ref{fig:ablation_mix} shows how audio semantic mixing is reflected in the generated visuals. We observe that our model successfully incubates multi-label audio mixtures and translates their combination in the resulting image in a rather coherent manner. It is worth emphasizing that some of these combinations may not be (or poorly) represented in the training data. For example, ``cows + police sirens'' or ``dog + fire crackling'' may be of rare if not no precedence in our training data. 

\begin{figure}
\vspace{-2mm}
    \centering
        \includegraphics[width=1.0\linewidth]{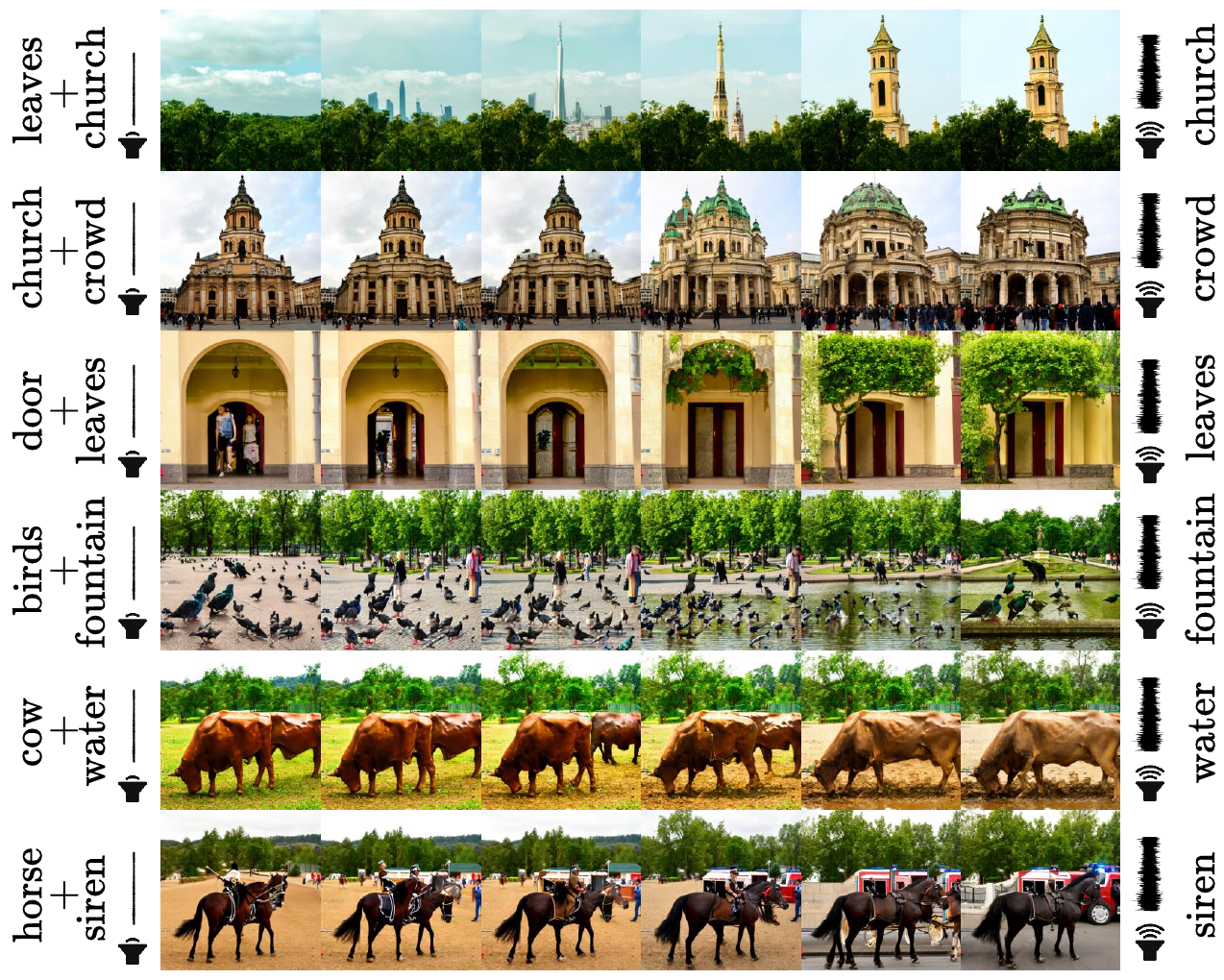}
        \vspace{-3mm}
    \caption{Impact of loudness variation for individual audio sources in the mix. From left to right, we increase the loudness for one of the sources while keeping the other one's fixed.}
    \label{fig:ablation_morph}
\vspace{-3mm}
\end{figure}

\noindent\textbf{Semantic Interpolation } Another important audio characteristic pertains to the notion of loudness. We aim at investigating our model's response to variations in loudness attributed to different semantic elements present in the mix. By systematically adjusting the loudness levels of each source prior to their combination, we seek to determine how these variations influence the model's interpretative process. Although no explicit considerations were brought to the inter-source loudness during our audio-visual pairing process\footnote{mixing coefficients $\gamma_m$ are sampled from a uniformly-distributed LUFS~\cite{grimm2010lufs} range with very small variance}, we observe that our model still successfully learns some aspects pertaining to semantic weights as they relate to the signal amplitude in the audio domain. Figure~\ref{fig:ablation_morph} illustrates the impact on the image generation process of gradually varying the loudness for one source while keeping the other one's fixed. We observe that, although there is a clear transition going from one semantic to another, the transformation seems rather abrupt. This could be due to the model encountering very small variance in the source loudness during training, leading it to focus on modeling whether an object is present in the image or not, without accounting for intermediate states. With additional explicit semantic weight modeling (\textit{e.g.} through depth maps), we expect the model to exhibit smoother interpolation, ultimately adding finer and greater level of control to the generation process. 

\begin{figure}
    \centering
        \includegraphics[width=1.0\linewidth]{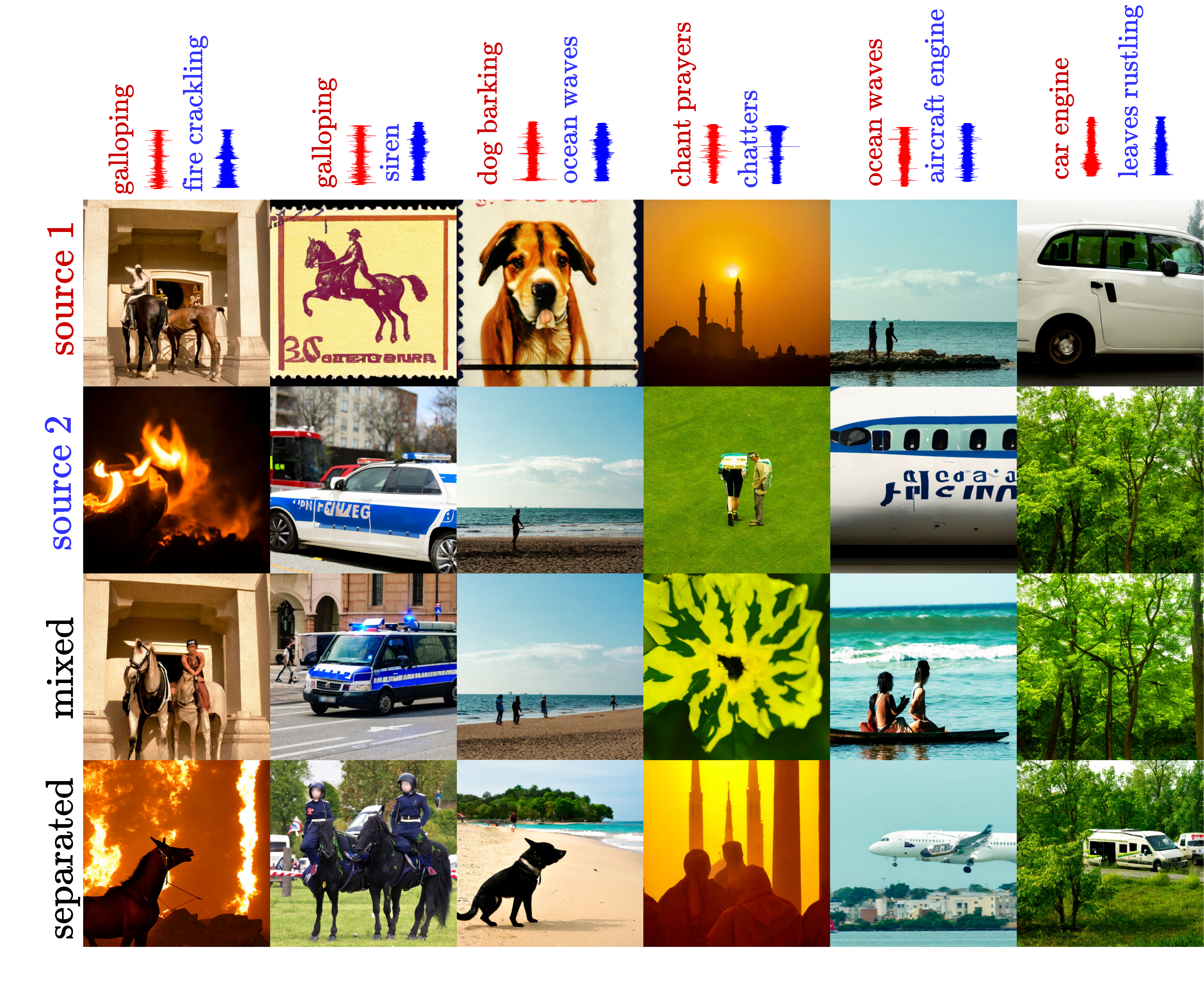}
    \vspace{-3mm}
    \caption{Impact of semantic source separation on generation.}
    \label{fig:ablation_sep}
\vspace{-4mm}
\end{figure}

 \noindent\textbf{Separated vs. Mixed } While our method shows robust behavior towards challenging and complex multi-semantic audio mixtures, in some cases overlapping sources occurring concurrently can lead the model to solely focus on the dominant source while ignoring the rest of the mixture. One potential solution to this failure case is to expose the model to individual sources separately instead of concurrently for instance by using source separation strategies. In this work, although we have complete access to the isolated sources during training, we always expose the model to audio mixtures rather than separated/concatenated sources. We here show the potential of applying source separation to real-world mixtures where isolated sources are not necessarily accessible. From Figure~\ref{fig:ablation_sep}, we observe that the model expectedly makes better sense of the input sonic prompt when the sources are not entangled (superposition in time) but rather individually presented one after the other. We hypothesize that the use of audio embedding backbones, which are more source-aware, would help in this case.

\begin{figure}
    \centering
        \includegraphics[width=1.0\linewidth]{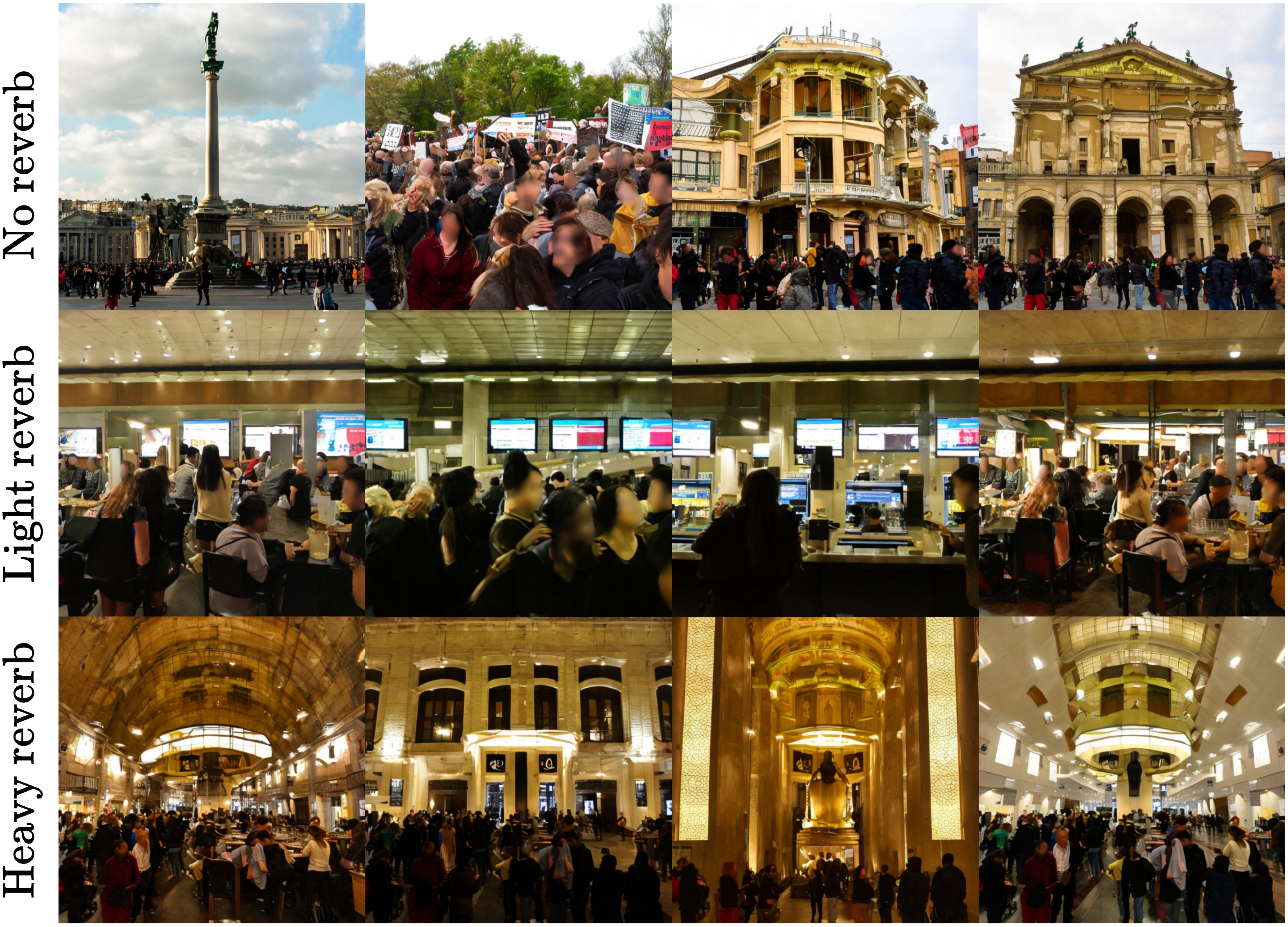}
        \vspace{-3mm}
    \caption{Chatter recorded in various acoustical places and their resulting visual counterpart. 
    }
    \label{fig:ablation_ir}
\vspace{-3mm}
\end{figure}

 \noindent\textbf{Learning Acoustics } Foleys include properties such as reverberation which reflect the acoustic response shaped by the physical space, causing a sound source to be heard differently in diverse acoustic environments (\textit{e.g.} church vs. bedroom). We aim to assess our model's ability to capture and translate these subtle acoustic characteristics in its image generation process. 
 By selecting recordings of crowd speaking in diverse acoustical spaces such as outdoor setting, offices, and churches, we evaluate the efficacy of our model to discern and visually represent these nuanced acoustic cues. In crafting audio prompts, we ensure the exclusion of any auditory cues other than speech to guarantee that the model solely utilizes the reverberations in visualizing the space. 
 Figure~\ref{fig:ablation_ir} shows the output of our model when being exposed to crowd noises in different acoustical settings. We observe robust integration of complex audio features into the visual domain, thereby enhancing the realism and contextual relevance of the generated images past the simple sound source themselves. We emphasize that no explicit consideration has been brought to the acoustic space modeling when assembling our training data hence the emergence of such a capability is hypothesized to be inherited indirectly from CLAP~\cite{clap2023wu} through the means of the retrieval process detailed in Sec.~\ref{sec:method}.

%% file: sec/6_conclusions.tex
\section{Conclusions}
In this work, we hypothesized that relying on in-the-wild videos to assemble semantically correspondent audio-visual pairs is overemphasized and it is evident from the prior literature that such a practice imposes severe restrictions on scale and diversity, hence impacting the ultimate practicality of the curated data. We argued that harnessing the remarkable multi-modal capabilities of today's VLMs and pretrained joint embedding feature spaces enable us to synthesize audio-visual pairs from disjoint uni-modal data sources, effectively sonifying any image-only dataset. We then demonstrated how one can utilize such a data to train competitively performant audio-to-image generative models. Finally, through a series of ablation studies, we examined a variety of interesting auditory capabilities that such models have implicitly acquired towards conditioning the image generation process.

%% file: sec/7_appendix.tex
\clearpage
\appendix
\section{Appendix}
\label{sec:appendix}
This Appendix section is organized as follows. We first present some image sonification examples to better showcase the dataset that we will publicly release. We then present additional generation instances from our models. We offer details on our data standardization process, including additional image filtering process and audio normalization. Lastly we provide insights on contrastively pre-training the audio projector, a technique that has been highlighted as essential in several prior studies. All images included in the figures of this section feature an ID (top left) to encourage readers to check their corresponding sonic counterpart, which will all be showcased on the dedicated project page.

\subsection{Dataset Overview}
\label{subsec:appendix_datasetoverview}

\begin{figure}[p]
    \centering
       \makebox[\textwidth][c]{ \includegraphics[width=0.8\textwidth]{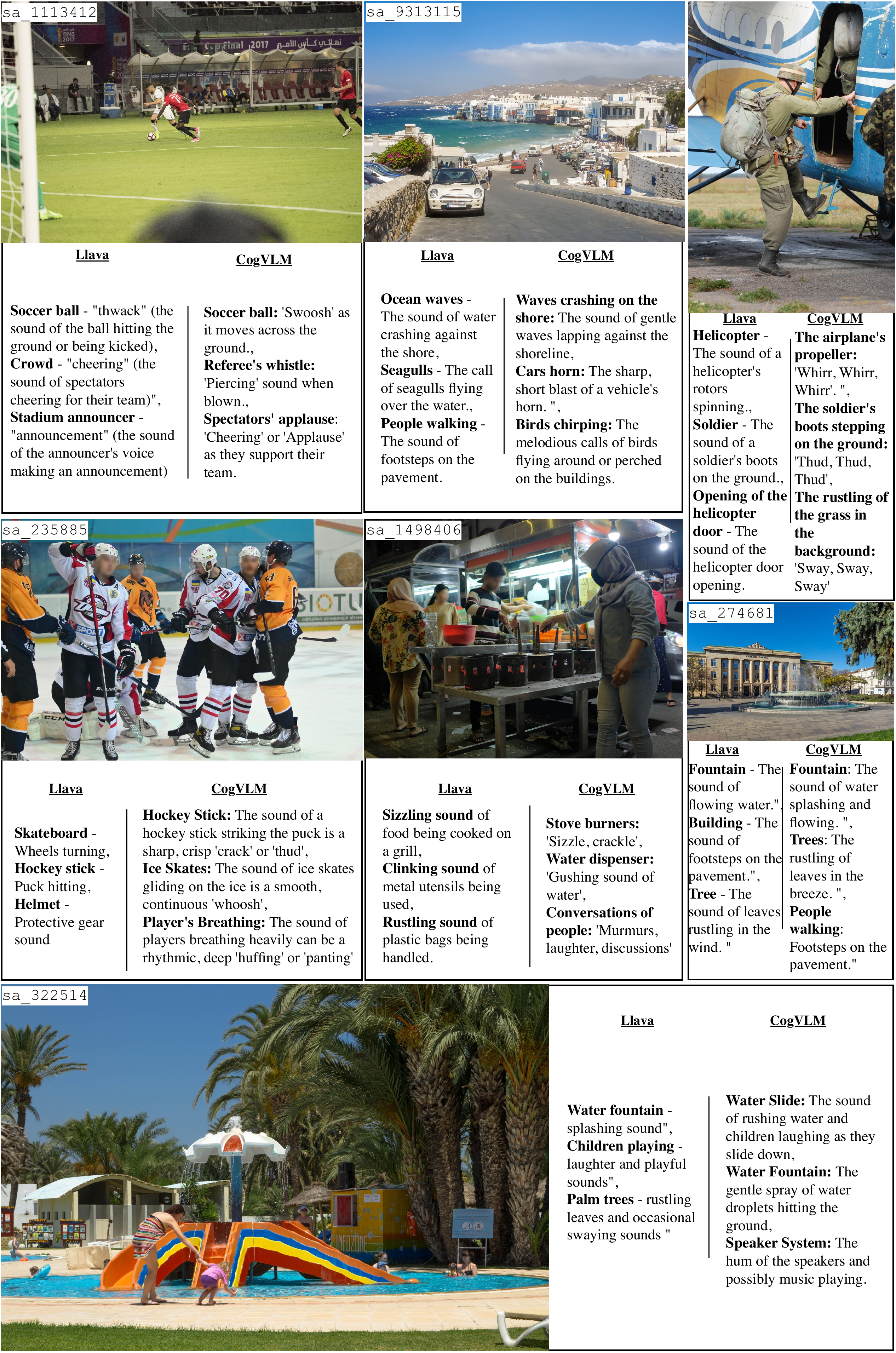}
       }
    \parbox{\textwidth}{\caption{Examples of sounding concepts inferred from the SAM dataset \cite{kirillov2023segany} using both Llava~\cite{liu2023llava} and CogVLM~\cite{wang2023cogvlm}.}
    \label{fig:concepts1}
    }
\end{figure}

\begin{figure}[p]
    \centering
       \makebox[\textwidth][c]{ \includegraphics[width=0.8\textwidth]{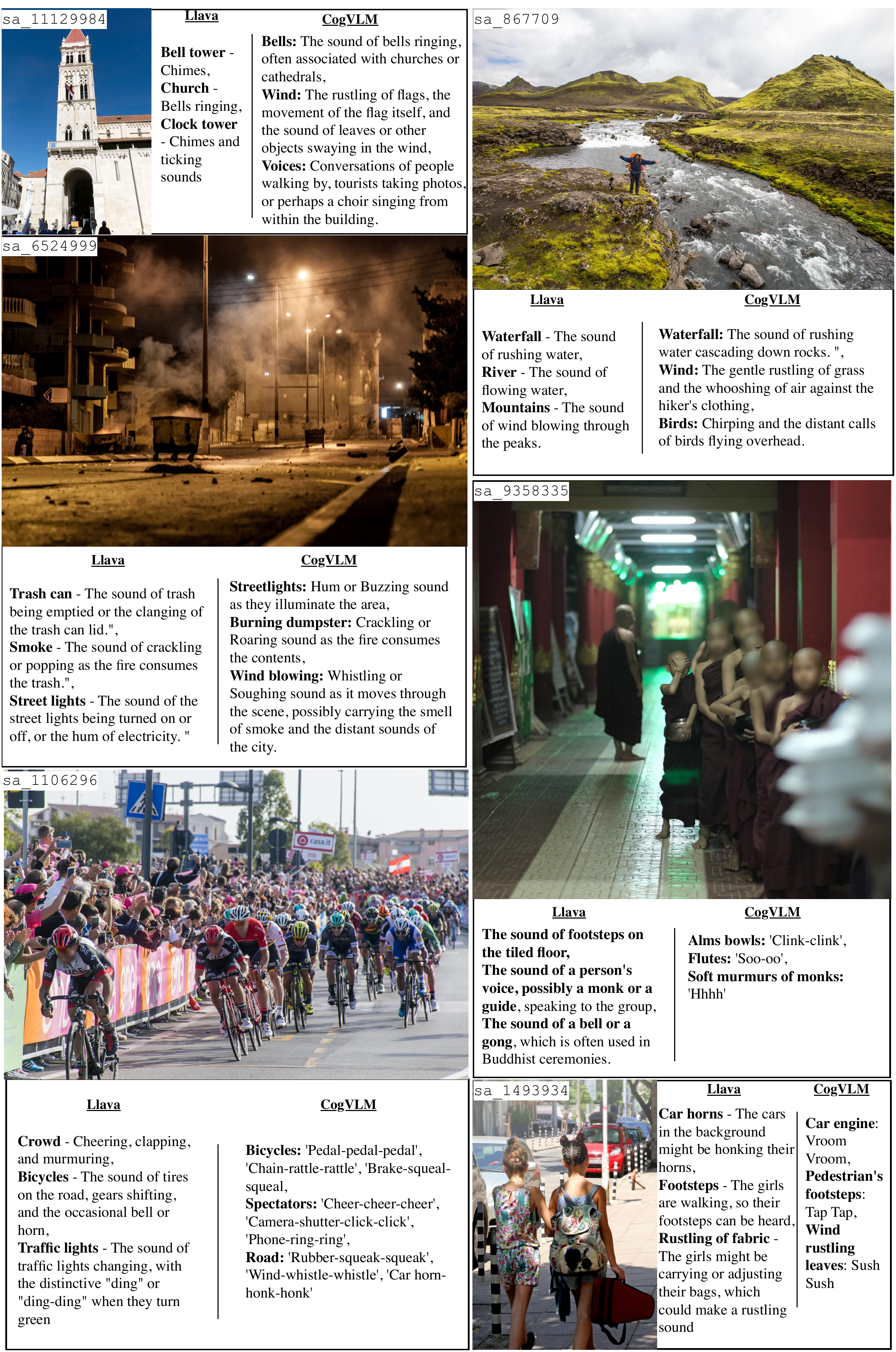}}
    \parbox{\textwidth}{\caption{Additional examples of sounding concepts inferred from the SAM dataset \cite{kirillov2023segany} using both Llava~\cite{liu2023llava} and CogVLM~\cite{wang2023cogvlm}.}
    \label{fig:concepts2}
    }
\end{figure}

One of the main contributions of this work is publicly sharing the code and model weights, in addition to our audio-visual dataset consisting of approximately 1M images, paired with textual sounding concepts and corresponding audio counterparts. The appeal of our data modeling approach lies in its modularity and flexibility. While we provide sounding concepts for a certain subset of SAM~\cite{kirillov2023segany} and our sound pool is composed of specific datasets, the various building blocks can effortlessly be extended at will without requiring major modification to our sonification pipeline. For instance, the pool of audio can be extended with additional audio samples to accommodate any new visual data and the potential new domain this data may bring in.
The sounding concepts will aim at issuing a comprehensive set of information needed in order to build and reproduce the audio-visual data showcased in this work. That is, for each image in our SAM subset, a set of entries, each denoting a sound file tied to a sounding concept is given. Along the file ID, we include additional information such as LUFS \cite{grimm2010lufs} normalization values, and confidence matching scores (\textit{i.e.} cosine similarity score). Ultimately, our sonification script (available upon code release) utilizes this metadata in order to produce the final audio counterpart of the images. Figures~\ref{fig:concepts1} and ~\ref{fig:concepts2} show examples of our dataset.

\subsection{Filtering}
\label{subsec:appendix_filtering}

\begin{figure}[h]
    \centering
        \includegraphics[width=1.0\linewidth]{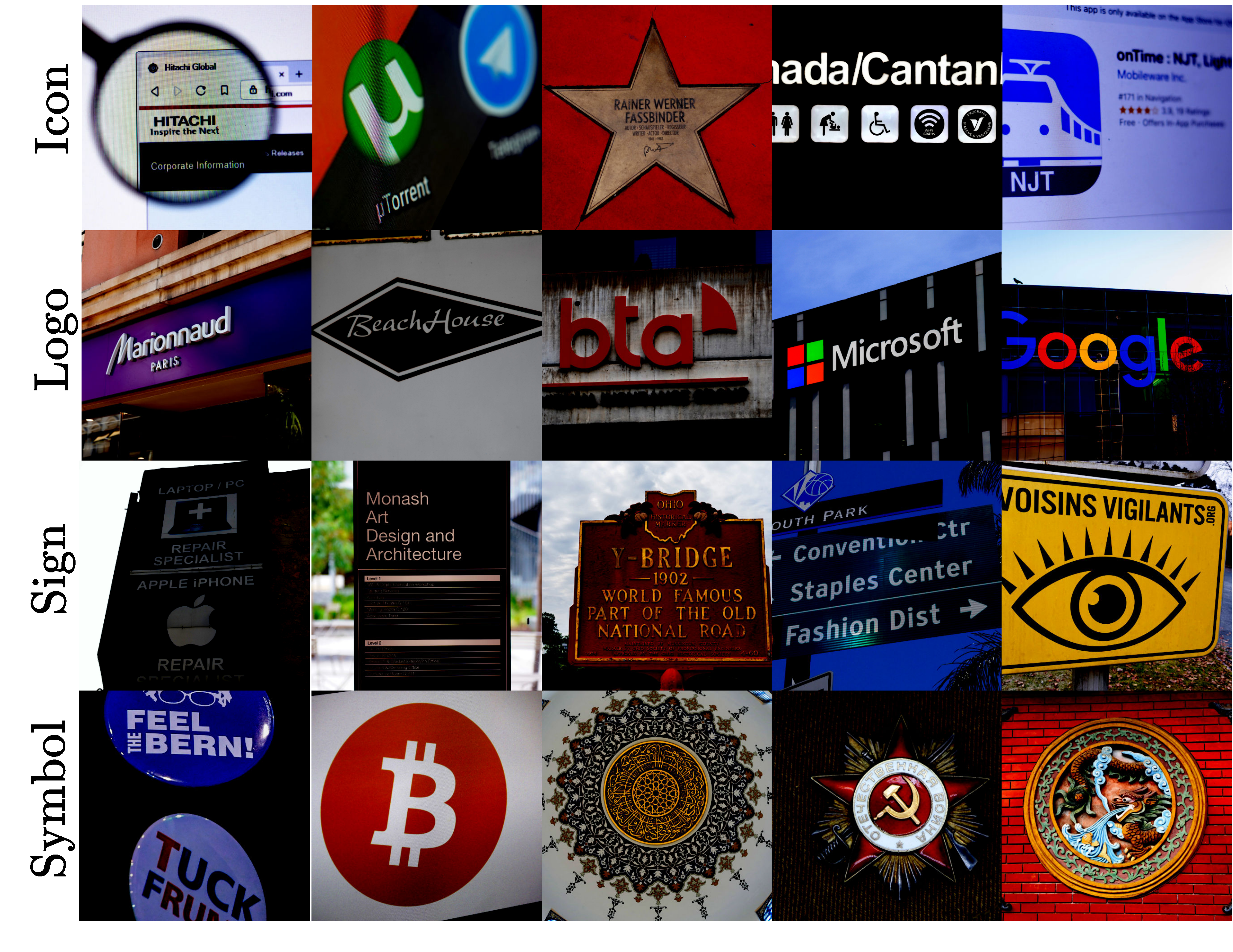}
    \caption{Examples of inherently silent images that are excluded from our dataset.}
    \label{fig:filtered}
\end{figure}

Many images in SAM~\cite{kirillov2023segany} do not carry any sonic characteristic and including them in our pool would hinder the training process. That is, images containing objects which are inherently silent are discarded from our training data altogether. We found that images depicting logos or symbols to be particularly detrimental to our task given that these type of images are ubiquitous in the SAM dataset. We therefore look for the following keywords to be present in the image captions as a condition for removal: \texttt{logo,icon,emblem,symbol,trademark,sign}. Namely we discard any image whose caption, previously obtained via Llava \cite{liu2023llava}, contains any of the above keywords. Figure~\ref{fig:filtered} shows some instances of images discarded from our training data using this process.

\subsection{More Qualitative Results}
\label{subsec:appendix_moreresults}

\begin{figure}[p]
    \centering
    \makebox[\textwidth][c]{
        \includegraphics[width=1.0\textwidth]{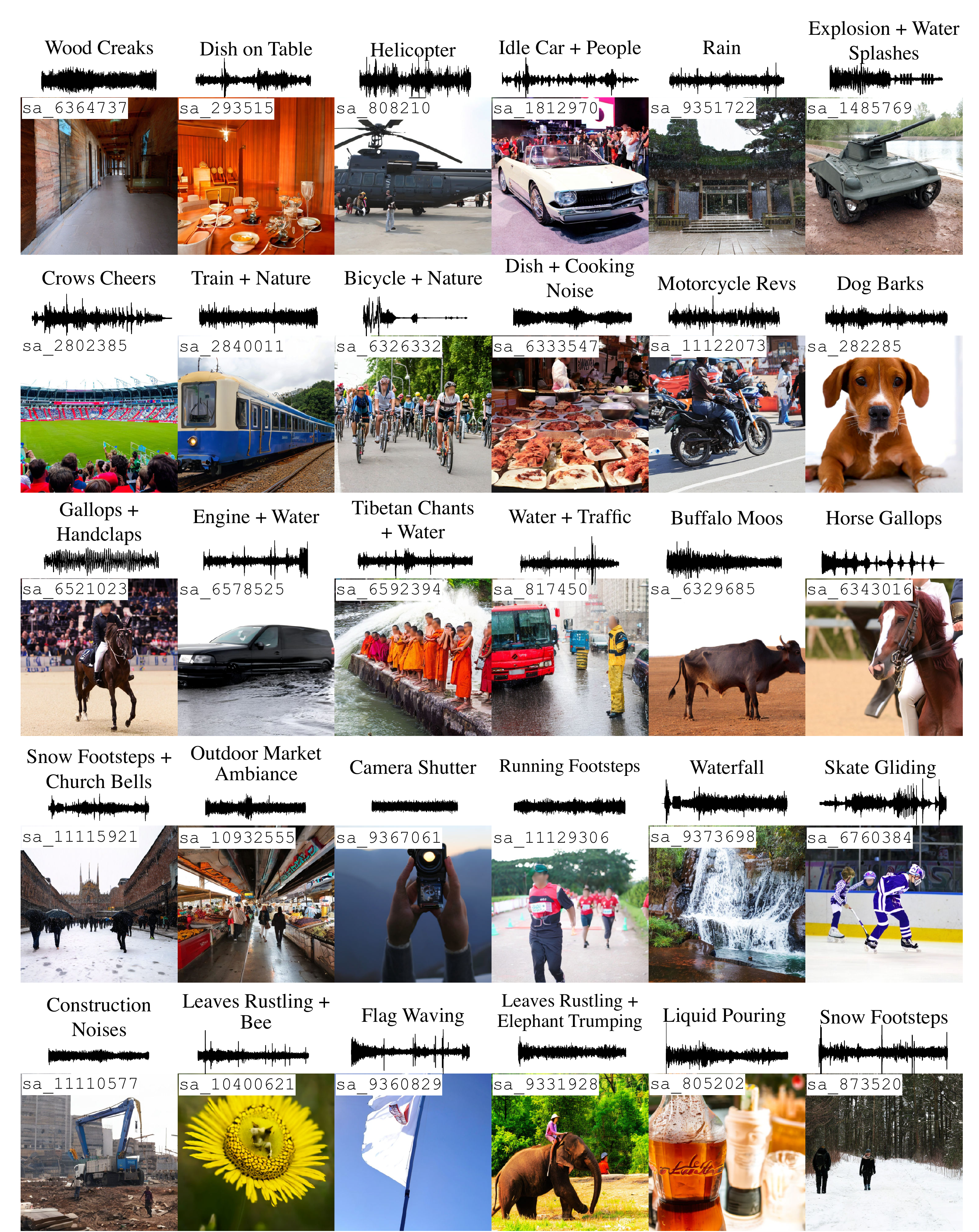}}
    \parbox{\textwidth}{\caption{More generated images by our model trained on our proposed audio-visual dataset. Each of the examples comes with a file ID (top left) to encourage readers to check their corresponding sonic counterpart on the dedicated project page. The descriptive text at the top of each example denotes a short keyword summary of what the input audio depicts.}
    \label{fig:more_outputs}
    }
\end{figure}

Figure~\ref{fig:more_outputs} depicts additional results by our audio-to-image generative model. In this case the model ingests in-domain audio samples drawn, and sometime mixed together, from our pool of audio samples. In order to build these validation audio mixtures, we held-out a small portion of the SAM images which he did not use during training.

\clearpage

\clearpage

\clearpage
\subsection{Failure Cases}
\label{subsec:appendix_failures}

Although our proposed approach has shown great promises and capacity towards robust and accurate image generation, it does come with its set of limitations and challenges. By following a thorough empirical qualitative assessment of the model's output, we compiled a list of most recurring failures cases that we have observed, most of which are addressable directly from the data modeling stage.

\begin{itemize}
\item \textbf{Blurry Faces:} This constraint inherently comes with the SAM dataset which blurs any human faces present in images. Our model trained on SAM images therefore intrinsically ingests the same blurring paradigm. A possible remedy is to do a light finetuning on another high-resolution dataset without blurred faces.
\item \textbf{Inference by Semantic Association:} One recurring challenge when prompting VLMs for sounding-concepts is to prevent it from assuming or being \emph{too creative}. Namely, we've observed many times that while the VLM would surely find sounding concepts appropriately matching the prompted image, sometimes the concepts could pertain to an object which is not actually present in the image. These concepts most often relate to background ambiance or atmosphere tightly connected with the semantics of the image. For example, images of beaches would often be associated with the sound of ``Seagulls'' even when no bird is actually shown.
\item \textbf{Drawing-Like Generations:} Another challenge induced by the choice of SAM as pool of training images is the heavy portrayal of art-like visuals \textit{e.g.} drawings and paintings. When prompted with such an image, the VLM will still returns plausible sounding concepts matching what the fictional image represents. Depending on the application, this may not be seen as an actual limitation, we however still deem relevant to mention it. 
\item \textbf{Statues:} Lastly, tightly bound to the limitation portrayed by drawings, statues are another type of inanimate/unsounding objects ubiquitously represented in SAM and yet treated as sounding object by the VLM. Note that while a "horse" could make sound, a statue of a "horse" cannot. This is a distinction the VLM has a hard time to make. 
\end{itemize}

\begin{figure}
    \centering
        \includegraphics[width=1.0\linewidth]{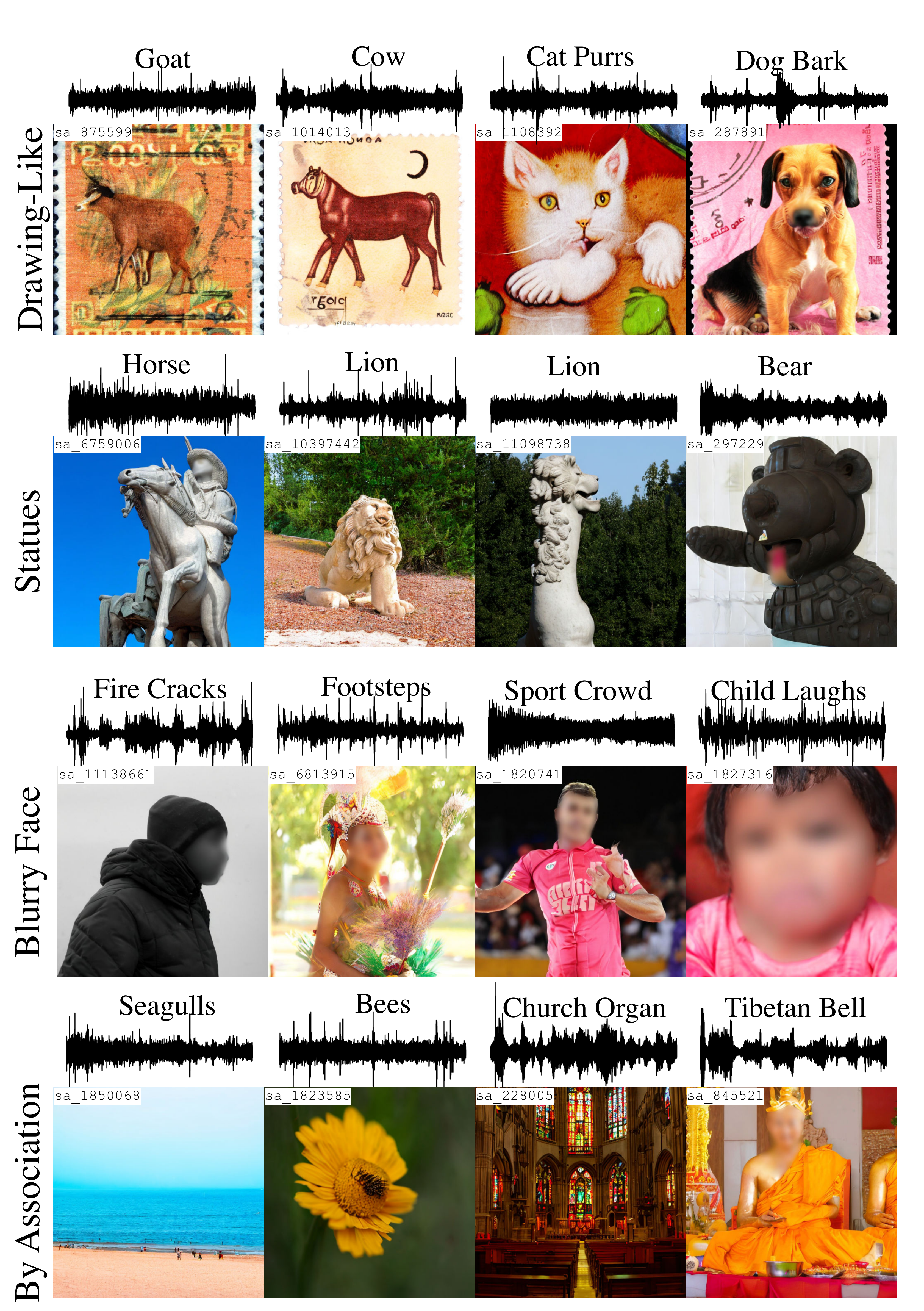}
    \caption{Limitations of our audio-to-image generation model that are rooted in the data curation process.}
    \label{fig:failure_cases}
\end{figure}

Figure~\ref{fig:failure_cases} showcases examples of the aforementioned limitations.

\subsection{Audio-visual dataset limitations}
\label{subsec:appendix_limitations}

\begin{figure}[h]
    \centering
        \includegraphics[width=1.0\linewidth]{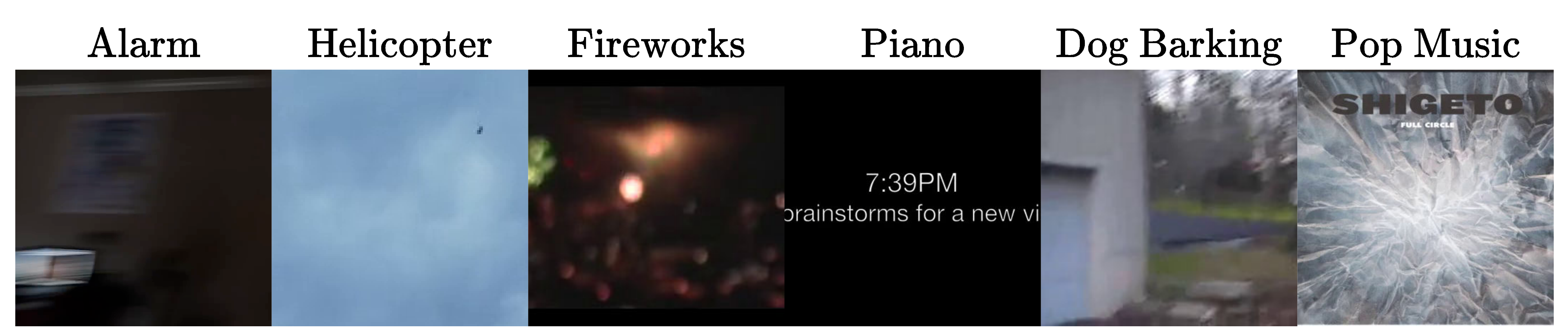}
    \caption{Examples of visual limitations induced by current audio-visual datasets, most often depicting low cross-modality correspondence.}
    \label{fig:limitations}
\end{figure}

``VGGSound'' \cite{chen2020vggsound} was proposed to partially address the shortcomings induced by Audioset \cite{gemmeke2017audioset} such as noisy audio, low-resolution video, and low cross-modality correspondence. Audio-visual pairs are typically extracted from such data by selecting one single frame, typically the center one, from the video, while keeping the entirety of the audio track. This comes without any assurance that the extracted image will indeed be of any relevance or correspondence to its sonic counterpart (\textit{i.e.} a dog can still be heard barking although it is temporarily out of the frame). Figure~\ref{fig:limitations} shows examples of such scenarios extracted directly from the VGG dataset. 

\subsection{Overfitting of existing models}
\label{subsec:appendix_overfitting}

\begin{figure}
    \centering
        \includegraphics[width=1.0\linewidth]{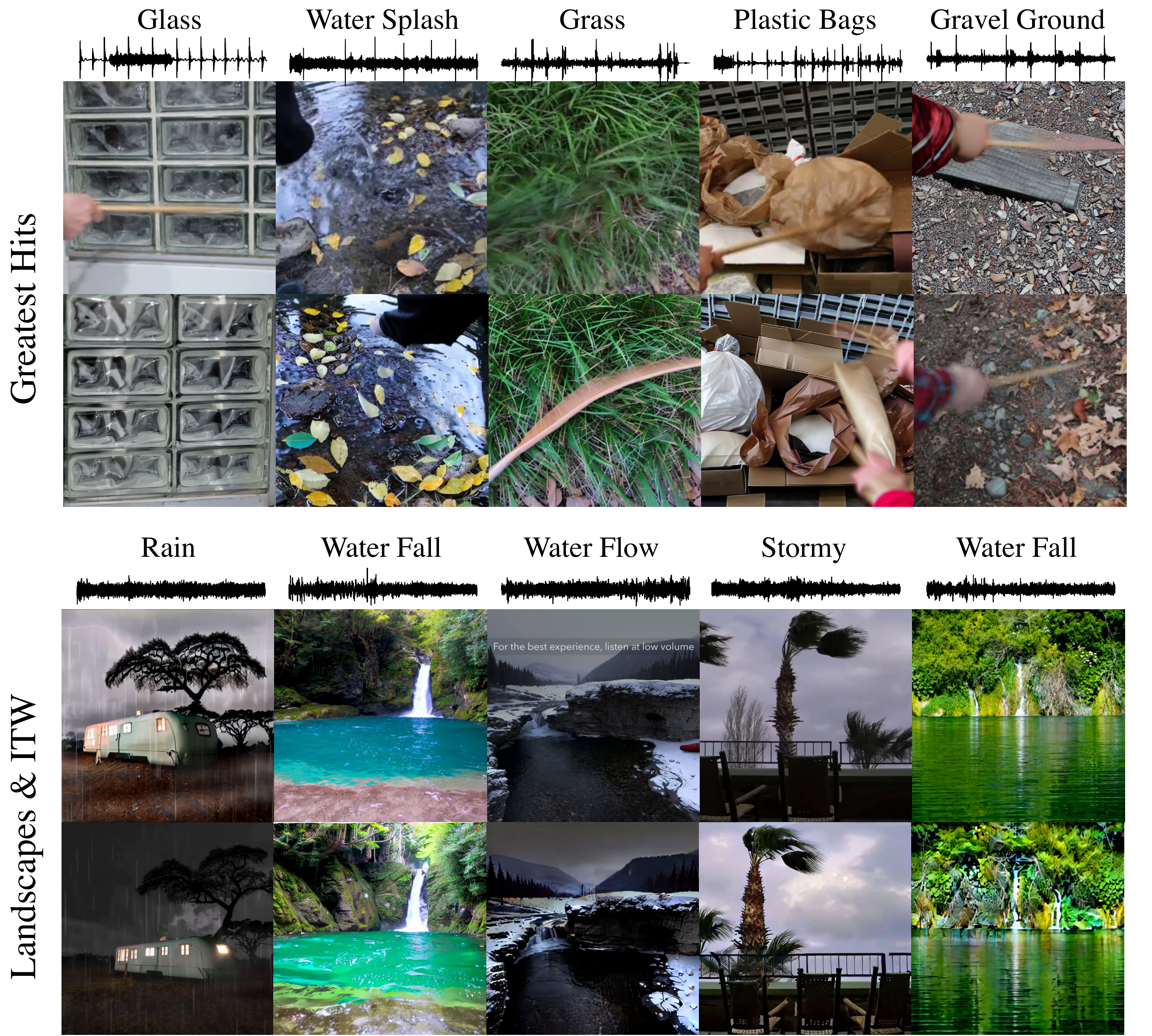}
    \caption{Some generated examples of SonicDiffusion~\cite{biner2024sonicdiffusion} overfitting on its training data. For reasons highlighted in Sec.~\ref{sec:results}, the same data is included as part of the models' evaluation, which can give them an unfair advantage. Top row denotes ground-truth image and bottom generated by SonicDiffusion.}
    \label{fig:overfitt}
\end{figure}

Our evaluation pipeline incorporates the entirety of the data for each of the datasets we have assessed (see Sec.~\ref{sec:results}), therefore some models have already encountered the test instances, and they may have over-fitted on the training data as well. Figure~\ref{fig:overfitt} presents evidence that some of these models, such as SonicDiffusion~\cite{biner2024sonicdiffusion} have indeed memorized their training data.

\subsection{Standardization Details}
\label{subsec:appendix_standardization}

To follow-up on Sec.~\ref{subsec:dataset}, we provide more details on the data curation process leading to our pool of audio examples that are ultimately exploited towards the sonification process. Audioset\cite{gemmeke2017audioset}, BBC sound effects (BBC SFX)\footnote{BBC sound effects library \url{https://sound-effects.bbcrewind.co.uk/}}, and FSD50k \cite{fonseca2022FSD50K} collectively total in 70k audio clips of varying lengths, sampling-rates, and channel configurations. To fulfill the requirements imposed by our downstream audio embedding stage, we standardize the audio files across all three datasets in the following manner; we resample the clips to a common 16kHz sampling-rate, sum any stereophonic files down to mono, and lastly segment them into 5-seconds long non-overlapping clips. For audio clips shorter than 5-seconds (e.g. such as frequently found in the FSD50K dataset) we zero-pad them to the target length. For instance, a 34 seconds long audio file undergoing this pre-processing stage would result in seven 5-seconds long non-overlapping chunks, the last one being trailed with 1-second worth of padding. Following this procedure, our audio pool totals in $~500,000$ 5-seconds audio chunks.

\subsection{Loudness Retention in AST}
\label{subsec:appendix_astloudness}

A key property of the audio embeddings, which we initially cover in Sec.~\ref{sec:audio_representation} pertains to the loudness information retention (\textit{i.e.} how much of the original signal amplitude is retained as part of the audio embeddings.). As we aim to maximize the degree of generative controllability through an effective and meaningful representation, we argue such property is crucially important. This type of information is intrinsically bound to the nature of the pre-trained model, more specifically what task and what type of data the model has been trained on. We therefore consider the following models as part of our proposed method; Hidden-Unit BERT (HuBERT) \cite{hsu2021hubert}, a self-supervised speech representation model providing rich and meaningful audio representations without the need for labeled data, and the Audio Spectrogram Transformer (AST) \cite{gong2021ast}, a state-of-the-art attention-based model for audio classification. AST is pre-trained on partially in-domain data (\textit{i.e.} Audioset) which makes it convenient towards our downstream application.
, HuBERT is on the other hand not and requires additional fine-tuning on Audioset. In the early stages of our experiments, we observed that AST yields more satisfactory results compared to Hubert hence we excluded Hubert from our later experiments to focus solely on AST.

\begin{figure}
    \centering
        \includegraphics[width=1.0\linewidth]{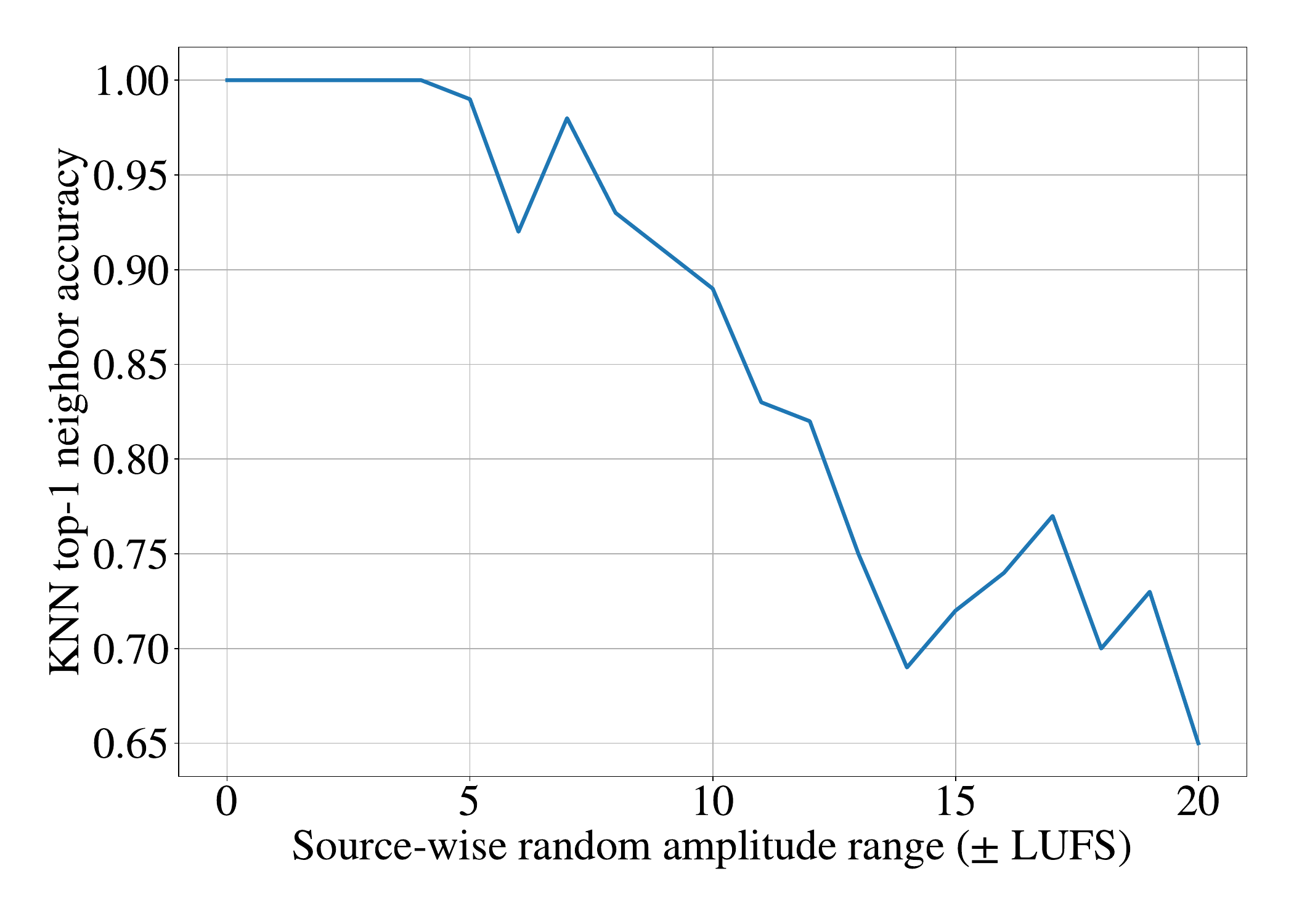}
    \caption{We perform KNN on AST embeddings and demonstrate how the classification accuracy evolves when each audio example is adjusted away from its original signal loudness, drifting away from their original embeddings.}
    \label{fig:ast_loudness}
\end{figure}

A priori, it is not clear how well the notion of signal loudness is retained within the AST embeddings. Hence, an experiment was crafted in which we gather a small set of audio examples encoded by AST model, each denoting their own ``class'', and proceeded to classifying each one of the examples using K-Nearest-Neighbor (KNN) algorithm over multiple iterations. At each iterations and prior to the AST embedding stage, the loudness of the audio signal is progressively increased. Intuitively and if indeed loudness impacts the embedding space, the accuracy of our KNN classifier should progressively deteriorate as the range of the loudness increases in the signal domain. Fig.~\ref{fig:ast_loudness} shows that the loudness information applied in the audio domain indeed has some impact on the AST embedding space, suggesting that some loudness information is still retained in the latent space.

\subsection{decibel-LUFS Normalization}
\label{subsec:appendix_lufs}

In this work we opt to normalize our audio signals following the deciBel-LUFS (Loudness Units Full Scale) \cite{grimm2010lufs} scale. This scale can be quite convenient when dealing with audio signals and human perception as it aims at measuring the perceived loudness of a given signal, in contrast to relying solely on the raw signal amplitude. For instance, two signals portraying identical peak level values may have totally different frequency responses (\textit{i.e.} bird singing vs. lion roaring), which dB-LUFS takes into account during its normalization process. Since, like any deciBel scales, LUFS is a relative scale, all of its values are negatives (\textit{i.e.} lower means quieter), 0.0-dB-LUFS representing the reference and maximal value. Next, we walk through a step-by-step process of normalizing an audio signal $x$ given a normalizing dB-LUFS value $t_x$. 

As the first step towards normalizing $x$, its integrated loudness needs to be computed. There are many loudness measurement algorithms available, each putting different emphasis and consideration on human perception. We opt for \cite{grimm2010lufs}, which models the human auditory perception and applies a weighting to the signal, accounting for frequency response and sensitivity. This process allows us to obtain a single loudness value $l_x$ pertaining to our current signal. This allows us to compute the difference in dB-LUFS between the current loudness $l_x$ and the target one $t_x$: $\text{G}=t_x - l_x$, namely the gain we will have to apply to $x$. Once $G$ is obtained, we can use it towards normalizing our signal $x$ in linear-scale fashion:

\begin{equation}
    x_\text{norm} = x \times 10 ^{\text{G}/20}
\end{equation}

\subsection{Implementation Details}
\label{subsec:appendix_implementation}

As in \cite{chen2023pixartalpha}, we compute the latent representation of $\mathcal{I}$ using a pre-trained variational autoencoder (VAE) from LDM \cite{rombach2021highresolution}. All images $i_m$ are first resized and crop-centered prior to the VAE input. We use a pre-trained DiT-XL/2 $512 \times 512$ model \cite{chen2023pixartalpha} as our base architecture. We point out that this model instance was originally pre-trained on visual representations beyond realistic photographs \cite{pan2023journeydb}. All of its weights are used towards the initialization of our audio-to-image model with the exception of the initial audio projector module, which we train from scratch instead. Our audio projector $\mathbb{A}$, analogous to the text projector $\mathbb{T}$ in the original model, consists of an multilayer-perceptron (MLP) which takes our audio mixture embeddings
as input to project it into vector representations with a hidden dimension resulting from the MLP projection, which is then ingested by the subsequent cross-attention blocks. In our early investigation, we experimented with various pre-training paradigms for $\mathbb{A}$, including contrastively learning a projection close to $\mathbb{T}$ \cite{biner2024sonicdiffusion}. However, we did not observe a substantial difference other than a marginal faster convergence of our model.

\subsection{Audio Projector Pre-training}
\label{subsec:appendix_projector}

Existing research \cite{yariv2023audiotoken, biner2024sonicdiffusion} on audio-to-image generation treats contrastively pre-training the audio projector with text representations as a crucial component of their training framework. This process aligns the audio projection with the text projection, making the two closer in a pseudo-shared embedding space. The rationale is that by projecting the pre-self-attention audio representation into a space that is similar to the one used by the text-to-image model, the modality transition becomes more seamless. This smoother shift can facilitate the training process, potentially leading to faster convergence and improved audio-to-image generation performance when model weights are initialized from a pre-trained text-to-image model. 

To investigate whether a contrastive pre-training stage is indeed a necessity or any beneficial to our framework we led an experiment in which we trained two model variants, each for a limited number of epochs (given limited compute resources); one with the audio projector being trained from scratch, while the other incorporating an audio projector contrastively pre-trained with text embeddings. For the latter, pre-training stage follows the same approach as in \cite{biner2024sonicdiffusion} where our training data consists of audio-text pairs.

\begin{figure}
    \centering
        \includegraphics[width=1.0\linewidth]{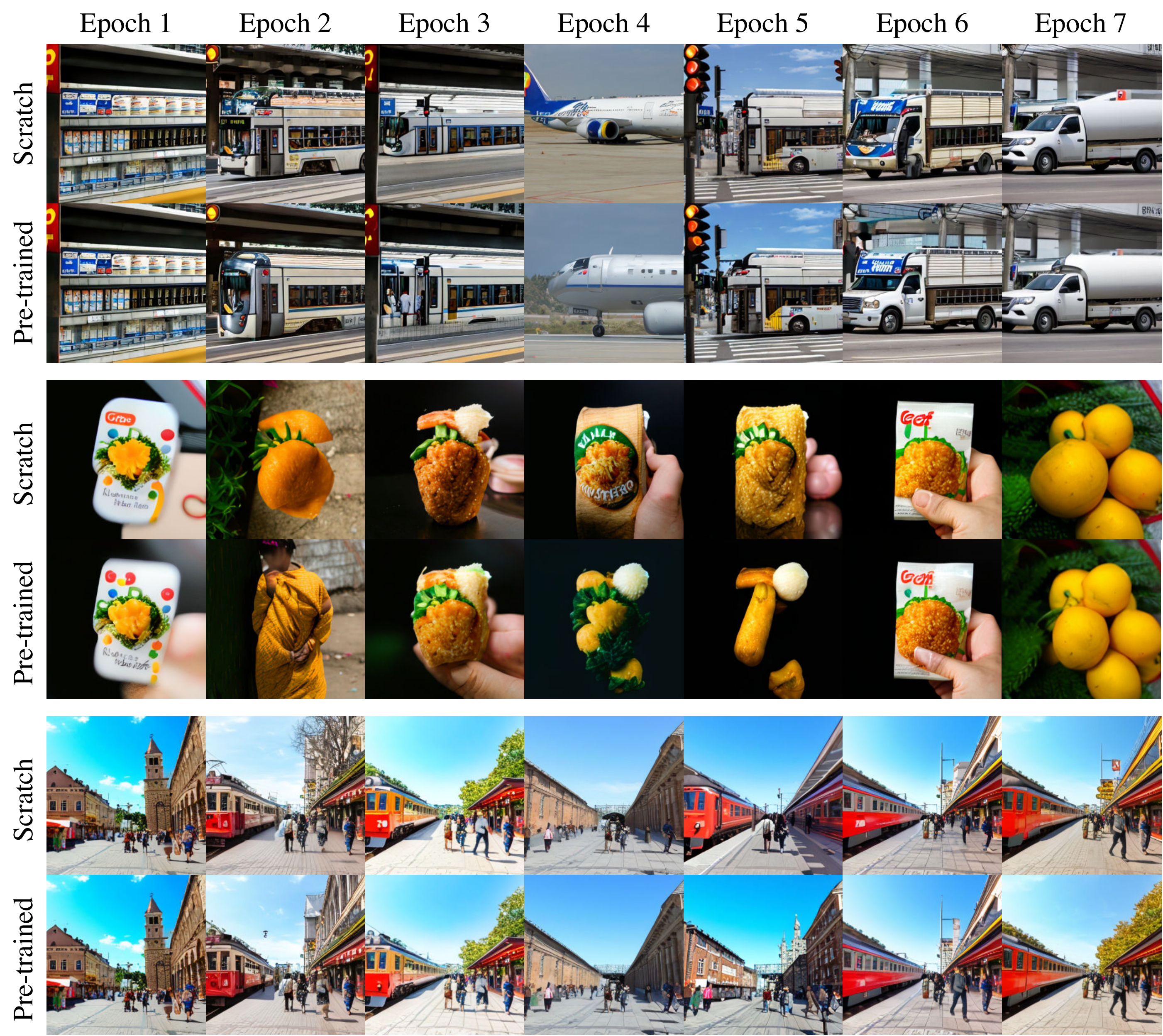}
    \caption{Comparative output examples over seven epochs between two model variants; one incorporating an audio projector that has been contrastively pre-trained using text embeddings (``Pre-trained'') and the other which is trained from scratch (``Scratch'').}
    \label{fig:projector}
\end{figure}

Figure~\ref{fig:projector} shows some qualitative results between two models demonstrating that there is little difference between the generation quality and pace of convergence. That is, unlike prior works, we avoid aligning audio embedding with text ones and train the former from scratch along side other layers of the diffusion backbone. 

